# Ion Pair Free Energy Surface as a Probe of Ionic Liquid Structure

Kalil Bernardino,[1*] Kateryna Goloviznina,[2]

Agílio A. H. Padua,[2] Margarida Costa Gomes,[2] Mauro C. C. Ribeiro[1]

[1] *Laboratório de Espectroscopia Molecular*, *Departamento de Química Fundamental*,

*Instituto de Química, Universidade de São Paulo, Av. Prof. Lineu Prestes 748, 05508-000, Brazil*

[2] *Univ Lyon, Ens de Lyon, CNRS UMR 5182, Université Claude Bernard Lyon 1,*

*Laboratoire de Chimie, F69342, Lyon, France*

*email: kalil.bernardino@gmail.com

The numerous combinations of cations and anions turn out possible to produce ionic liquids with fine-tuned properties once the correlation with the molecular structure is known. In this sense, computer simulations are useful tools to explain and even to predict properties of ionic liquids. However, quantum mechanical methods are usually restricted to either small clusters or short timescales, so that the use of parametrized force fields is needed for studying the bulk liquids. A method is proposed in this work to enable a comparison between quantum mechanical and both polarizable and non-polarizable force fields by means of the calculation of free energy surfaces for the translational motion of the anion around the cation in gas phase. This method was tested for imidazolium-based cations with 3 different anions, $[BF_4]^-$, $[N(CN)_2]^-$, and $[NTf_2]^-$. It was found better agreement with the DFT calculations when polarizability is introduced in the forcefield. In addition, the ion pair free energy surfaces reproduced the main structural patterns observed in the first coordination shell in molecular dynamics simulations of the bulk liquid, proving to be useful probes for the liquid phase structure that can be computed with higher level methods and the comparison with forcefields can indicate further improvements in their parametrization.

## I.    INTRODUCTION

Ionic liquids (ILs) are unique substances with unusually low melting points when compared with common salts while also having very low vapor pressure. The vast number of possible combinations of cations and anions enables the production of ILs with a wide range of physical properties that can be fine-tuned if the correlation with the chemical structure is known. Since the first development of water stable room temperature ILs, computer simulations have been used for this task, being able not only to explain but also often to predict their properties before experiments.[1] Molecular dynamics simulations of ionic liquids are normally performed with fully-atomistic force fields[2-4] and help to explain, at a molecular level, physical properties like the viscosity,[5,6] the melting point,[7] dielectric response,[8,9] sound modes,[10] and bulk and shear moduli[11] of ILs. Traditional non-polarizable force fields keep the partial charges on the atoms fixed, with polarization represented implicitly by an effective Lennard-Jones or similar potential. Despite a good description of the structural and thermodynamic properties, the fixed integer-charge models predict too slow dynamics when compared to experiment[12] and fail in reproducing dielectric constants.[8,9] During the last decade this problem was avoided by scaling down the net charge of the ions by a factor of *ca.* 0.7–0.8,[13,14] which allowed to accelerate the dynamics by reducing the Coulomb interactions. However, this procedure results in decreased densities up to 5%,[15,16] underestimation of cohesive energy, incorrect dipole distributions,[17] and poor description of ILs in mixtures.[12]

These issues can be remediated by including polarization explicitly in a more realistic force field. Polarizable force fields give a good prediction of dynamic properties, such as viscosity and diffusion coefficients, without weakening near-neighbour structure.[8,17] Explicit polarization leads to strengthening of the first-neighbour ionic shell due to an increase in short-range electrostatic interactions and disordering of the second-neighbour shell due to the weaker long-range correlations when compared to fixed-charge models. Recent polarizable force fields for ionic liquids include APPLE&P by Borodin[18] based on induced point dipoles, the *ab initio* force field by Yethiray *et al.*[19] and CL&Pol by Padua *et al*[16] based on Drude induced dipoles. Among these, CL&Pol can readily be used for major families of ionic liquids (and possibly extended to new ones) and is compatible with commonly used force fields for molecular compounds and materials, such as OPLS-AA.[20]

Despite the great improvements over ILs properties prediction by including polarization effects, a comparison with *ab initio* calculations is desirable in order to validate the polarizable model for the structural description of the condensed phase. However, due to the large number of particles and slow relaxation of ILs, even with state of art computational resources the use of quantum mechanical approaches is limited to small clusters or bulk simulations with production times of only a few picoseconds.[21] Despite being very low, the vapor pressure of ionic liquids is not zero, and mass spectroscopy measurements showed that the most common species in ILs vapor were neutral ion pairs with no free ions detected close to the room temperature,[22] and the calculations of these ion pairs are feasible with standard quantum mechanical methods. However, the main interest regarding these systems lays in the liquid phase rather than the gas phase studies. On one hand, the liquid properties will depend on many body interactions as each ion is surrounded by several others instead of pairing with a single opposite charge one. On the other hand, some properties showed remarkable correlations with ion pairs formed in the condensed phase, for instance, the viscosity that have a linear relation with the ion pair lifetime,[5] indicating that some extrapolations from gas phase to the liquid phase structure may be possible.

In this work, we present a method to probe the structure in the first coordination shell of the liquid by means of calculations of an ensemble of ion pair structures. The results are shown as free energy surfaces for the translational motion of the anion around the cation and enable a quantitative comparison between the results obtained by quantum mechanical methods with the ones obtained by both polarizable and non-polarizable forcefields for the gas phase ion pairs. These surfaces are in agreement with similar free energy surfaces computed in liquid phase by means of molecular dynamics simulations, demonstrating that the ion pair sampling is a suitable way to use high level computational methods to extrapolate structural information for ILs and also to evaluate the accuracy of the force field employed in condensed phase studies. A comparison with a popular semi-empirical method was also included to evaluate the description of a lower cost quantum mechanical method.

## II. METHODS

### Ion pair free energy surface calculation

Three ionic liquids were studied by means of molecular dynamics simulations of the liquid phase and free energy calculations of the ion pair in the gas phase: 1-butyl-3-methylimidazolium tetrafluorborate, $[C_4C_1Im][BF_4]$, 1-ethyl-3-methylimidazolium dicyanamide, $[C_2C_1Im][N(CN)_2]$, and 1-ethyl-3-methylimidazolium bis(trifluoromethanesulfonyl)imide, $[C_2C_1Im][NTf_2]$. Figure 1 shows the molecular structures of these ions, in which the names assigned to some atoms are given to be used in the data analysis. This selection aims an increasing complexity regarding the anion, from the rigid and highly symmetric $[BF_4]^-$, passing by the rigid and planar $[N(CN)_2]^-$, to the flexible and low-symmetry $[NTf_2]^-$. In particular, the flexibility of $[NTf_2]^-$ will introduce additional concerns when comparing the predictions between the structure of gas phase ion pair and in the liquid phase. The methodology used in the study of the gas phase ion pairs will be presented in this section while the one for the liquid phase will be discussed in the following section.

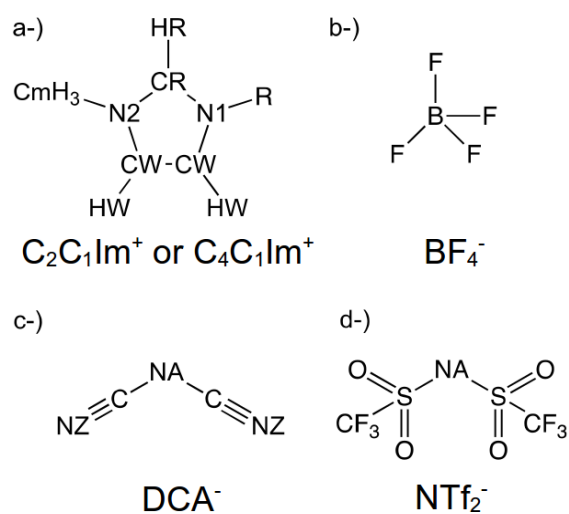

**Figure 1.** Structural formulas and abbreviations of the ions studied in the present work with the names attributed to atoms used in specific analyses. a-) 1-butyl-3-methylimidazolium (R = $C_4H_9$) or 1-ethyl-3-methylimidazolium (R = $C_2H_5$) cation, and the anions b-) tetrafluorborate, c-) dicyanamide, d-) bis(trifluoromethanesulfonyl)imide.

In order to study the distribution of the anion around the cation for gas phase ion pair, spherical free energy surfaces were computed around selected atoms of the cation. A spherical grid with 92 points was defined for the translational ($t$) movements of the anion around the N2 atom (see Figure 2 for a representation and Figure 1 for atoms naming), being the central atom of the anion (B for [BF$_4$]$^-$, NA for [N(CN)$_2$]$^-$ and [NTf$_2$]$^-$) set up over these points. For each translational point, 42 rigid body rotations ($r$) were performed with the anion and, for each rotation, 8 precessions ($p$) were performed for [N(CN)$_2$]$^-$ and [NTf$_2$]$^-$ and 5 precessions for [BF$_4$]$^-$. This procedure results in a total of 30912 structures for each surface for [N(CN)$_2$]$^-$ and [NTf$_2$]$^-$ and 19320 for [BF$_4$]$^-$. The points which would generate strong repulsive interactions due to atom superpositions were excluded from the calculations since its contribution to the summation in Equation (1) would be negligible. This procedure excluded approximately half of the generated structures. The energy of remained structures, $E(t,p,r)$, was computed using a single point quantum mechanical or force field method. The relative free energy of the grid point $t$ around the cation is calculated by the well-known equation for the Helmholtz free energy in a canonical ensemble:

$$\Delta F(t) = -k_B T \ln\left[\sum_{p,r} \exp\left(-E(t,p,r)/k_B T\right)\right] - F(t\,')$$

(Equation 1)

where $k_B$ is the Boltzmann constant, $T$ is the temperature, and $F(t')$ is the smallest free energy obtained in the grid. The same methodology was applied by Colombari *et al.* to search for the most favorable adsorption sites for DNA basis over a nanoparticle,[23] and the software developed by Colombari, THEMIS, was used to produce the structures for the ion pair and to calculate the free energy according to Eq. (1).

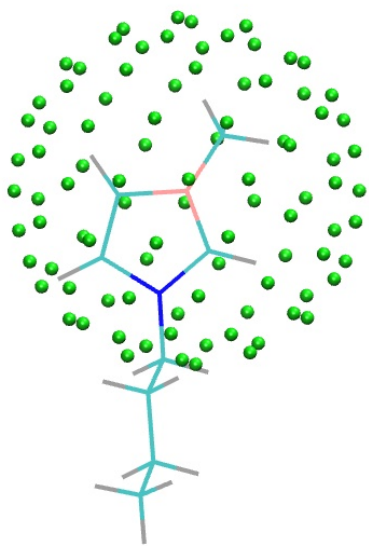

**Figure 2.** The 92 points grid equidistant from the N2 atom (highlighted in pink) of $[C_4C_1Im]^+$ cation. An ensemble of structures was produced by positioning the central atom of the anion over each one of these points. At every point, rotations and precessions of the anion were performed as discussed in the text. The energy was calculated for each structure using several methods in order to obtain the free energy surface for the ion pair according to Eq. (1).

The energies $E(t,p,r)$ for the ensemble of ion pair structures were computed using four different methods: *i.* a DTF method as the highest theory level, *ii.* the polarizable[16] and *iii.* the non-polarizable[2,3] version of CL&P force field, *iv.* a semi-empirical calculation in order to test the response of a low cost quantum mechanical method. These calculations were performed as following:

*i.* DFT calculations using the CAM-B3LYP functional[24] with basis set Def2-TVZPD[25] and D3BJ dispersion correction[26] done with Orca 4.0.0.2 package.[27] The basis set superposition error was computed for some randomly selected structures, but was observed that this error was small and approximately constant for a given ion pair at the considered distances, so that it was neglected since our interest relies in the energy differences of different configurations of an ion pair.

*ii.* Drude polarizable force field CL&Pol,[16,28] with the energy minimization of the Drude oscillators and the final energy obtained with the LAMMPs software[29] (see details about force fields in the next section).

*iii.* Non-polarizable force field CL&P,[2,3] for which the energies were computed directly by the Colombari software.

*iv.* Semi-empirical calculations with the PM7 functional done with the MOPAC2016.[30]

Despite the equilibrium geometry of the ions could be slightly different in each one of the compared methods, we focused on the interaction between the ions and how this affects the distribution of the anion around the cation. Hence, exactly the same structures were employed in each of those methods and the energy of isolated ions were computed and subtracted from the dimer energy in Equation 1 for each method. The equilibrium geometries of the ions were obtained from DFT calculations with the CAM-B3LYP/Def2-TZVPD method. No implicit solvation model was used in quantum mechanical calculations or forcefield calculations of ion pairs since one of the aims of this work was to compare how the structure of ion pairs in vapor phase are related to the structure in the liquid phase. Besides this, a good match was found between the descriptions in gas and in the liquid phase, and some differences between them are more likely to be related to other effects like molecular flexibility that would not be corrected by the inclusion of implicit solvation as is discussed in Results and Discussion section.

**Molecular dynamics simulations**

Molecular dynamics simulations were carried out using the LAMMPS program package.[29] Initial configurations of periodic cubic boxes containing 300 ion pairs of ionic liquid were produced using `fftool`[30] and `packmol`.[32] A cutoff of 12 Å was considered for the intermolecular interactions with tail corrections for energy and pressure applied. Bonds with hydrogen atoms were constrained with the SHAKE algorithm. Electrostatic energies were evaluated using the particle-particle particle-mash (PPPM) method with an accuracy of $1 \times 10^{-5}$. The timestep was of 1 fs.

We used the fixed-charge CL&P force field[2,3,33,34] with modified parameters for Lennard-Jones interactions involving fluorinated ions[35] and the polarizable CL&Pol force field[16] for the simulation of ionic liquids, with integer net charges in both cases.

The CL&Pol treats polarization explicitly through Drude induced dipoles.[27] The mass of Drude particles was set at 0.4 a.u. and the force constant of the Drude core–Drude particle bond was fixed at 4184 kJ mol$^{-1}$. The partial charges of Drude particles were calculated from $\alpha = q_D^2/k_D$ using atomic polarizabilities taken from the work of Schröder.[36] Drude dipoles were positioned on all heavy atoms; polarizability of hydrogen atoms was summed on the polarizability of atoms they are bounded to. We applied Thole damping function[37,38] to reduce short range electrostatic interaction between Drude dipoles with a universal parameter of $a = 2.6$. The motion of Drude particles with respect to the cores was thermostated at 1 K as an approximation of self-consistent regime.

The CL&Pol force field requires modification of the Lennard-Jones (LJ) parameters of the original non-polarizable model, in particular scaling $\varepsilon_{ij}$ between fragments, in order to exclude implicit polarization effects, since they are already accounted for explicitly by the Drude dipoles.[16,39] We considered the $[C_2C_1im]^+$ cation and the anions $[BF_4]^-$, $[N(CN)_2]^-$, and $[NTf_2]^-$ as entire fragments, while $[C_4C_1im]^+$ was split into a cation head group $[C_2C_1im]^+$ and a $C_4H_{10}$ alkyl chain.

Scaling factors of interactions between fragments $i$ and $j$, $k_{ij}$, were evaluated using symmetry adapted perturbation theory, SAPT (Equation 2),[40]

$$k_{ij} = \frac{E_{disp}}{E_{disp} + E_{ind}} \quad \text{(Equation 2)}$$

where $E_{disp}$ and $E_{ind}$ are, respectively, dispersion and induction contributions to the total energy. The scaling factora were taken from the works of Padua and are reported in Table I. A complete presentation of the CL&Pol forcefield together with the calculated physical properties is given in the work of Goloviznina *et al.*.[16]

**Table I.** Scaling factors for $\varepsilon_{ij}$ parameter in LJ potential obtained at the SAPT2+/aDZ level.

| Fragments | $k_{ij}$ |
|---|---|
| $[C_2C_1im]^+ \ldots [N(CN)_2]^-$ | 0.61 |
| $[C_2C_1im]^+ \ldots [NTf_2]^-$ | 0.65 |
| $[C_2C_1im]^+ \ldots [BF_4]^-$ | 0.52 |
| $[C_2C_1im]^+ \ldots C_4H_{10}$ | 0.76 |
| $C_4H_{10} \ldots [BF_4]^-$ | 0.51 |

Drude induced dipoles were added to the simulation setup using the `polarizer` tool and scaling of LJ interactions was done with `scaleLJ` script. The USER-DRUDE package was enabled in LAMMPS to handle Drude dipoles. Equilibrations of 2 ns were followed by 10 ns runs in the *NpT* ensemble using Nosé-Hoover thermostat and barostat in the temperature range of 303-343 K and at the pressure of 1 bar. TRAVIS[41] software was employed for analyses of the trajectories produced and VMD 1.9.3[42] was used to render the structure, spatial distribution functions and free energy surfaces representations.

## III. RESULTS AND DISCUSSIONS

### A. 1-butyl-3-methylimidazolium tetrafluorborate, [C₄C₁Im][BF₄].

Figure 3 shows the relative free energy surfaces for [$C_4C_1Im$][$BF_4$] ion pair at a fixed distance of 0.388 nm between N2 and F atoms. This distance corresponds to the equilibrium distance obtained by a full optimization of the ion pair with CAM-B3LYP/Def2-TZVPD. The surfaces computed for other distances, but still inside the first coordination shell, are given in Supporting Information (Figures S1 and S2). Two representations are given for each surface: at the top (**a**, **c**, **e,** and **g**)  with the imidazolium ring on the plane of the paper, and at the bottom (**b**, **d**, **f**, and **h**) with the imidazolium ring perpendicular to the paper with the HR atom pointing toward the reader. The red-most regions correspond to the smallest free energy values, thus being the regions with larger probability of finding the anion at the given spherical shell, while blue-most correspond to the highest free energy values (unfavorable regions for the anion). According to the scale at the bottom of Figure 3 (with values in kJ/mol) one should notice that even small color differences correspond to large free energy variations. The same kind of representation will be used for the other ionic liquids discussed in this paper (Figures 7 and 12).

In the case of [$C_4C_1Im$][$BF_4$], the dissemblance between the four methods tested are noticeable. In DFT calculation (Figure 3 **e** and **f**), which one expects to be the most reliable between the methods used, the most favorable region for the anion is close to the hydrogen HR, which is the most acid hydrogen of the imidazolium ring.[43] In the lowest energy structure for this surface, the distance between the HR and the closest F atom of the anion is 1.93 Å, which is within the limit usually considered as a hydrogen bond, with an angle CR-HR-F of 155º (see representations of the lowest energy structures found for each method at this spherical shell in Figure S14 of S.I.). In the case of the non-polarizable force field (Figure 3 **c** and **d**), the same region becomes unfavorable, indicating that the interaction between the HR and F atoms is probably underestimated by the force field, and the region above/below the ring are the most favorable ones. For the non-polarizable force field, there is no signal of a hydrogen bond like interaction between the ions in the lowest energy structure found, with the distance between the HR and the closest F atom being 2.8 Å and the angle CR-HR-F of 104º. Introduction of polarizability (Figure 3 **a** and **b**) partially corrects this failure and brings again the most favorable position close to the HR atom, but still it doesn't match exactly the DFT calculation and the lowest energy structure has HR-F distance of 2.17 Å and CR-HR-F angle of 115º. However, one should notice that the Drude polarizable force field works better than the semi-empirical PM7 functional for the region around the HR atom for this ion pair.

Variations between the methods are also noticeable for this ion pair in the region close to the $CH_3$ group bonded to imidazolium. In this region, both the polarizable and non-polarizable force fields overestimate repulsion comparing to the DFT calculation, however, the difference is smaller for the polarizable one. Similar tendencies were also noticed for the surfaces at other distances (see Supporting Information). These results indicate that a smaller σ parameter in the Lennard-Jones potential between the anion and the methyl or a softer potential for the repulsive part of the van der Waals interaction between these groups may improve further the structural description of this ionic liquid by the forcefield, as well as the use of different interaction parameters for the hydrogen atom bonded to the C2 atom of imidazolium ring that is so far described by the same non-bonded parameters as the other hydrogen of the ring.[2,3,16]

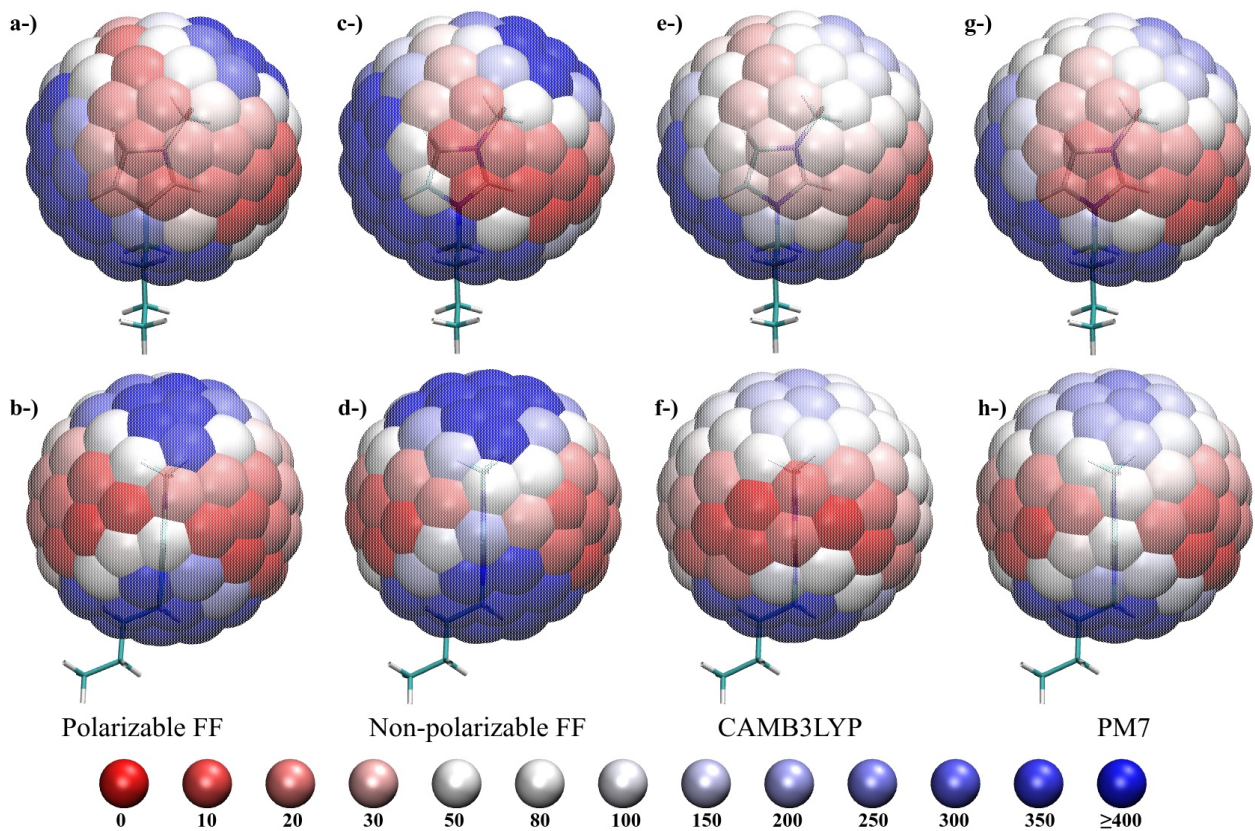

**Figure 3.** Free energy surfaces for the [C₄C₁Im][BF₄] ion pair with the distance N2-B of 0.388 nm. The colors corresponds to different free energy values according to the scale at the bottom with values in kJ/mol. **a** and **b**-) Polarizable force field; **c** and **d**-) non-polarizable force field; **e** and **f**-) DFT calculation; **g** and **h**-) semi-empirical calculation. The top representations are views from above the plane of the imidazolium ring. In the bottom representations, the HR atom of imidazolium points toward the reader. See Supporting Information for other distances.

The free energy surfaces enable one to quantify the most and the less favorable regions for the anion around the cation for the gas phase ion pair at different levels of theory. A natural question is how these differences affect the structure of the condensed phase, where a cation is not surrounded by one but by several anions and by other cations. Also, a limitation of the methodology used for the study of the free energy surfaces is that the geometries of the ions were kept rigid along the scan, while some deformations should be expected in relation to the equilibrium geometry of the isolated ions due to the interactions in both the ion pair and the liquid phase. In order to tackle these questions, the comparison with molecular dynamics simulations of the liquid was done using the polarizable and non-polarizable force fields. The radial distribution function, $g(r)$, for the boron atom around selected atoms of the cation is given in Figure 4, and for the fluorine atoms in relation to the HR atom is given in Figure S9. There is no significant change in the first coordination shell besides a slight tendency for the anion get closer to HR and CM (carbon from $CH_3$ bonded to the ring) atoms of the cation when polarizability is introduced. A reduction in the long-range structure is noticed by the reduction of the $g(r)$ peaks for the subsequent coordination shells when polarizability is explicitly included in the force field. This is due to the absence of dynamic screening in the non-polarizable forcefields that results in artificially enhanced of long-range correlations.[8,38]

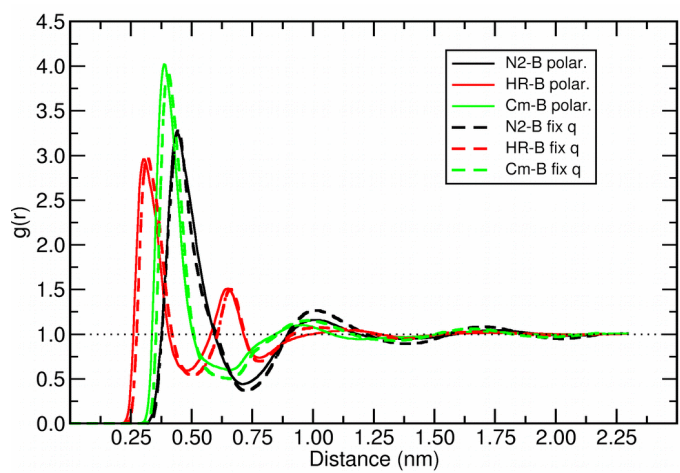

**Figure 4.** Radial distribution function, $g(r)$, for the B atom of $[BF_4]^-$ around selected atoms of $[C_4C_1Im]^+$ calculated from liquid simulations using the polarizable (solid lines) and non-polarizable (dashed lines) force fields.

The same density of the anion in the first shell around the HR in both force fields seems to contradict the differences observed in the free energy surfaces. However, the spatial distribution of the anion is different and that is not captured by the $g(r)$ analyses. This is noticeable in both the spatial distribution function (sdf, see Figure 5) and also in the distribution of the angle CR-HR-F for the fluorine atoms in the first shell around the HR atom (Figure S10). The CR-HR-F angle is shifted toward $180^{\circ}$ when polarizability is introduced as expected by the lowest energy structures for the gas phase ion pair. This shift demonstrates an improvement in relation to the non-polarizable force field, since it brings the angle closer to the most favorable value according to the DFT calculation. In the sdf, the increase of the concentration of the boron atom in the region close to HR in the plane of the ring is observed when polarizability is introduced (Figure 5 **b**) at the same time that the density above/below the ring is reduced (Figure 5 **a**). The same tendency is observed in the relative free energy surfaces (Figure 3 **a**, **b**, **c** and **d**): The relative free energy gets smaller in the region at the plane of the ring and close to HR for the polarizable force field. The changes in the free energy for the ion pair modify the local density in the liquid, but this happens in a way that the average density in a spherical shell around the HR atom does not change appreciably leading to almost the same $g(r)$ of the anion atoms around the HR (Figures 4 and S9).

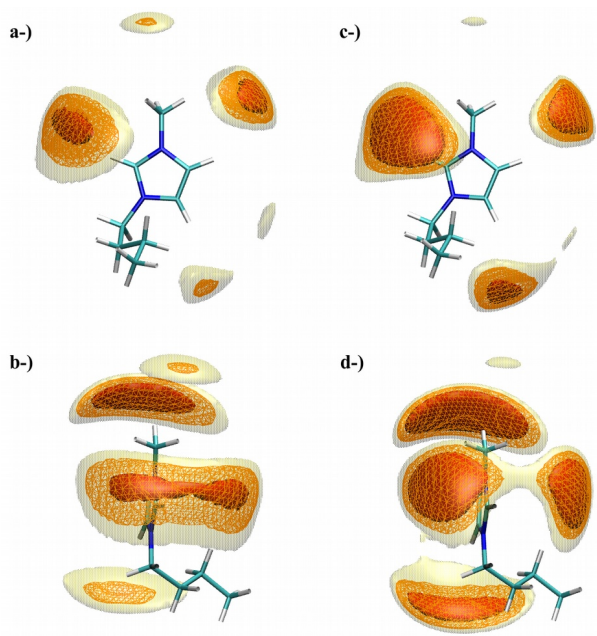

**Figure 5.** Spatial distribution function (sdf) of the B atom of $[BF_4]^-$ around $[C_4C_1Im]^+$. **a** and **b**: polarizable force field; **c** and **d**: non-polarizable force field. The top representations (**a** and **c**) are views from above the plane of the imidazolium ring. In the bottom representations (**b** and **d**), the HR atom of imidazolium points toward the reader. Isosurface values: 16 nm$^{-3}$ (translucid yellow), 20 nm$^{-3}$ (orange wireframe) and 26 nm$^{-3}$ (solid red). The difference between the distributions of polarizable and non-polarizable models is given in Figure S18.

Two more subtle changes can be noticed in the sdf: Increase of the anion density above the $CH_3$ bonded to the ring when introducing the polarizability (compare **a** and **c** in Figure 5), that is also correlated with a reduction of the repulsion in this region as observed in the ion pair (Figure 3), and a larger density in the region between the HW atoms and the butyl group in the non-polarizable model. The last effect was unnoticed for the ion pair because the radii of the free energy surfaces were not large enough to sample this region.

The differences between the sdfs for both force fields are seen more directly by subtracting the sdf of the non-polarizable force field from the one of the polarizable model. This difference is shown in Figure S18, where the blue surfaces correspond to the regions in which there is an increase in the anion average density in the polarizable model in relation to the non-polarizable one, and the red surfaces indicate reduction of the average density. Similar images are also provided in SI for the other ILs studied here.

A more direct comparison with the free energy surface of the gas phase ion pair can be made by computing a similar free energy surface from the sdf of Figure 5 for the liquid phase. The time-averaged density value $\rho$ in a volume element $V_i$ is related to the relative free energy of the anion according to Equation 3,

$$G_r\left(V_i\right) = -k_B T \ln\left(\frac{\rho_r\left(V_i\right)}{\rho_{r,max}}\right)$$ (Equation 3)

where the subscript $r$ indicates that only the volume elements distant from the N2 atom by the same radius are used in the corresponding gas phase surface (in practice, a finite thickness d$r$ of 0.015 nm was employed for the sampling in the liquid and the volume element between $r - \mathrm{d}r$ and $r + \mathrm{d}r$ was considered). The reference zero for the free energy values were defined in a similar way as for the gas phase surfaces: The smallest free energy found in the spherical shell was subtracted from every value computed. This can be done by dividing the densities in each point by the largest density found in the spherical shell, $\rho_{r,max}$.

The liquid phase free energy surfaces for the [BF$_4$]$^-$ shown in Figure 6 are in excellent agreement with the results obtained for the gas phase for both of the models with Drude oscillators and with fixed partial charges, especially considering the limitation of rigid molecules in the calculus for the gas phase ion pair and the fact that the grid definition is not totally equivalent in the two cases. Again one can see a reduction of the free energy of anion close to the methyl group of imidazolium for the polarizable model in comparison to the non-polarizable one (compare Figure 6 with Figure 3 **a**, **b**, **c** and **d**). The lowest free energy region (red-most region in Figures 3 and 6) is in both cases more concentrated above/below the imidazolium ring for the non-polarizable force field (representations **c** and **d**) while in the polarizable one (**a** and **b**) the region close to the HR atom and the plane of the ring becomes more favorable. In other words, in absence of polarizability, the free energy barrier to move between the two sides of imidazolium ring is excessively high. Due to the flexibility of the molecules in the molecular dynamics simulation and the interaction of each ion with several other cations and anions in the liquid phase, the well-defined regions observed for the gas phase ion pair are blurred out. Nevertheless, the essential features of the first coordination shell in condensed phase and the main differences between the two models were correctly captured by the free energy surfaces calculated for the gas phase ion pair.

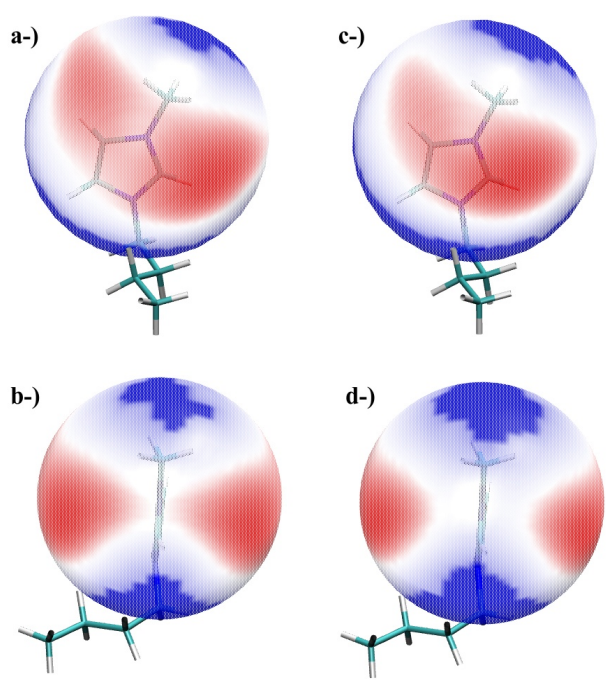

**Figure 6.** Relative free energy surface for the translation of $[BF_4]^-$ in a spherical shell of 0.388 nm around $[C_4C_1Im]^+$ computed in the liquid phase. **a** and **b**: polarizable forcefield; **c** and **d**: non-polarizable forcefield. The top representations (**a** and **c**) are views from above the plane of the imidazolium ring. In the bottom representations (**b** and **d**), the HR atom of imidazolium points

toward the reader. The red-most regions correspond to the smaller free energy values while the blue-most correspond to the largest free energy values of the spherical shell.

A higher free energy barrier for anion translation is expected to increase the viscosity of the IL, as observed by Zhang and Maginn.[7] Together with the artificial enhancement of the structure in second and further coordination shells,[8] the increase in the free energy barrier for the $[BF_4]^-$ movement between the most favorable positions (above and below the ring) contributes to the artificially high viscosity when using non-polarizable force fields. When introducing explicit polarizability, the computed viscosity of $[C_4C_1Im][BF_4]$ and the other ILs get closer to experimental values.[16]

## B. 1-ethyl-3-methylimidazolium dicyanamide, $[C_2C_1Im][N(CN)_2]$.

Differences between DFT, force fields and semi-empirical approaches were also found in the free energy surfaces for the $[C_2C_1Im][N(CN)_2]$ ion pair (see Figures 7 and S3), although of smaller magnitude than the ones observed in the previous case. As in the $[C_4C_1Im][BF_4]$ case, the non-polarizable force field (representations **c** and **d** in Figure 7) results in relatively larger free energies for the anion in the plane of the ring and close to the HR atom in comparison with both the DFT and the polarizable model, which implies in a smaller density of the anion in this region in the liquid phase. However, unlikely the case with $[BF_4]^-$, even in the DFT calculation this region is not the most favorable in the case of $[N(CN)_2]^-$, which prefers the region close to HR but above/below the ring instead of the region at the plane. Thus, in this case all the models agree qualitatively fot the most favorable region for the anion, but the force field without polarizability overestimates the difference between the most favorable and the other regions around the cation. Another variation arises, however, at the opposite side of the ring around the HW atoms, where both the forcefields and even the PM7 calculation result in smaller relative free energies than would be expected with bases in the DFT surface (this difference is more noticeable at larger N2-NA distance, see Figure S3). This discrepancy may need different interaction parameters for the non-equivalent hydrogen atoms of imidazolium ring to be developed in future parametrizations in order to be corrected. The four methods result in the same lowest energy structure shown in Figure S15 (some appears at opposite sides of the imidazolium ring, but they are equivalent due to the cation symmetry).

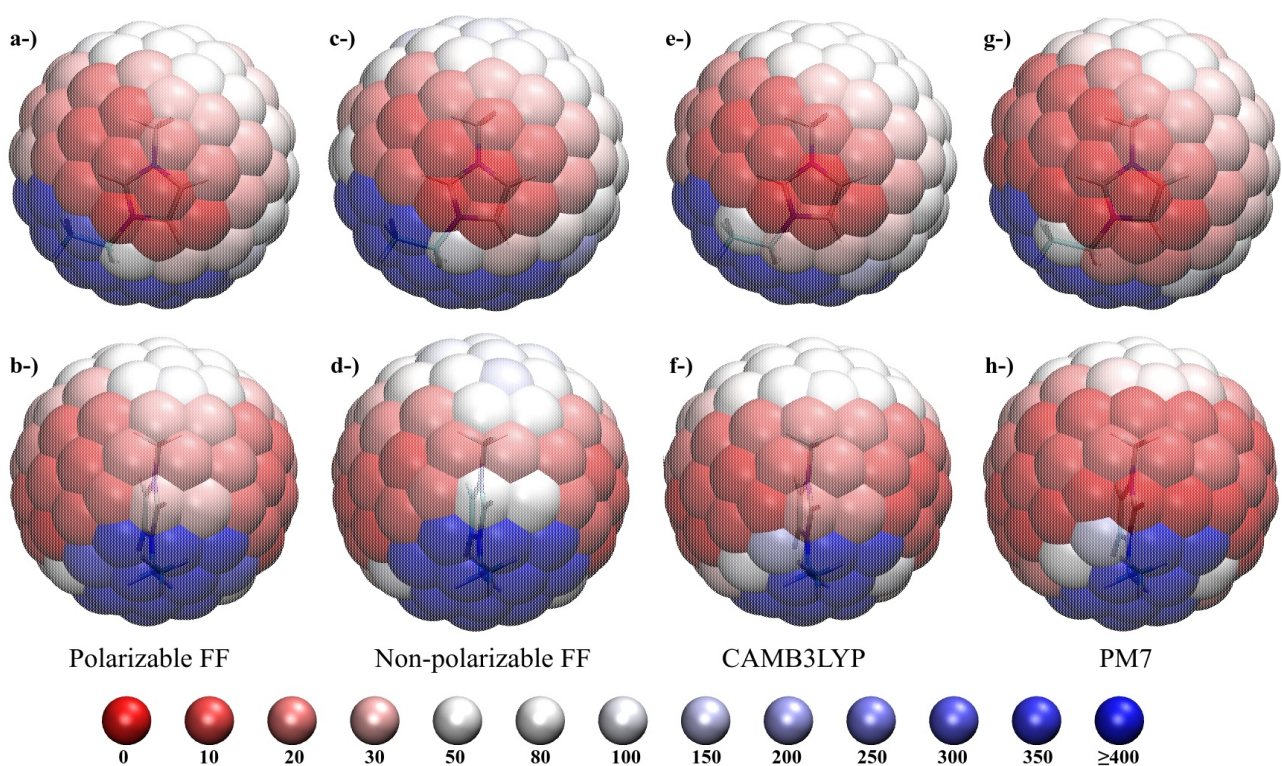

**Figure 7.** Free energy surfaces for the $[C_2C_1Im][N(CN)_2]$ ion pair with the distance N2-NA of 0.41 nm. The colors correspond to different free energy values according to the legend at the bottom (kJ/mol). See Supporting Information for other distances. **a** and **b**-) Polarizable force field; **c** and **d**-) non-polarizable force field; **e** and **f**-) DFT calculation; **g** and **h**-) semi-empirical calculation. The top representations are views from above the plane of the imidazolium ring. In the bottom representations, the HR atom of imidazolium points toward the reader.

In the case of $[C_2C_1Im][N(CN)_2]$, differences in the radial distribution functions of the liquid phase arise for the central nitrogen atom of the anion around selected atoms of the cation (Figure 7). For the non-polarizable force field (dashed lines), there is stronger correlation at short distances in the distributions involving the central atom of the anion, which is more noticeable for the distribution around the HR atom. This result can be compared with the same distributions for the terminal nitrogen atoms of the anion, NZ (see Figure S11). The distribution of NZ around HR exhibits the opposite tendency observed for the NA (red curves in both figures), with the correlation at short distances stronger for the polarizable force field (solid red lines) than for the non-polarizable one (dashed red lines). This indicates that in both force fields there is a tendency of the terminal nitrogen atoms pointing toward HR instead of the central nitrogen. This finding agrees with the lowest energy structures of the ion pair calculations (see Figure S15), though this tendency is stronger in the case of the polarizable forcefield. As discussed in the previous system, there is also a systematic reduction of the structure for the second and further coordination layers in the $g(r)$ due to the dynamic electrostatic screening in the polarizable force field.

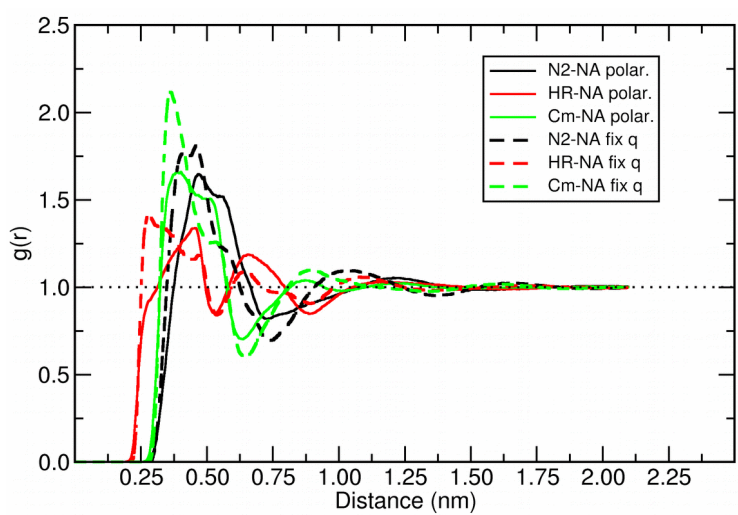

**Figure 8.** Radial distribution function, $g(r)$, for the NA atom of DCA$^-$ around selected atoms of $[C_2C_1Im]^+$ calculated from liquid simulations done with polarizable (solid lines) and non-polarizable (dashed lines) force fields.

A more detailed view of structural changes is provided by the spatial distribution function of the anion central atom around the cation shown in Figure 9 and by the plot of the difference between these distributions (Figure S19). Despite the region above/below the imidazolium ring being the most favorable in both cases, an increase of the density next to the HR atom in the same plane of the ring is observed when polarizability is included in the model (compare representations **b** and **d** in Figure 9 or see the subtraction of the surfaces in Figure S19). The isosurface of 9.5 nm$^{-3}$ and even the one of 8.5 nm$^{-3}$ only "connects" the both sides of the ring through the side with the HR atom in the polarizable model. This implies in a greater mobility of the anion between the two sides of the ring in the polarizable forcefield, what can contribute to the larger diffusion coefficients when polarizability is introduced alongside other effects such as the reduction in the long-range structure due to the electrostatic screening.[8] The reduction of the anion density in this region in the absence of polarizability may act qualitatively similar to the methylation of CR carbon, which reduces the diffusion coefficient and increases the viscosity, a fact that is interpreted as the loss of the anion ability to move between the two faces of the imidazolium ring.[7]

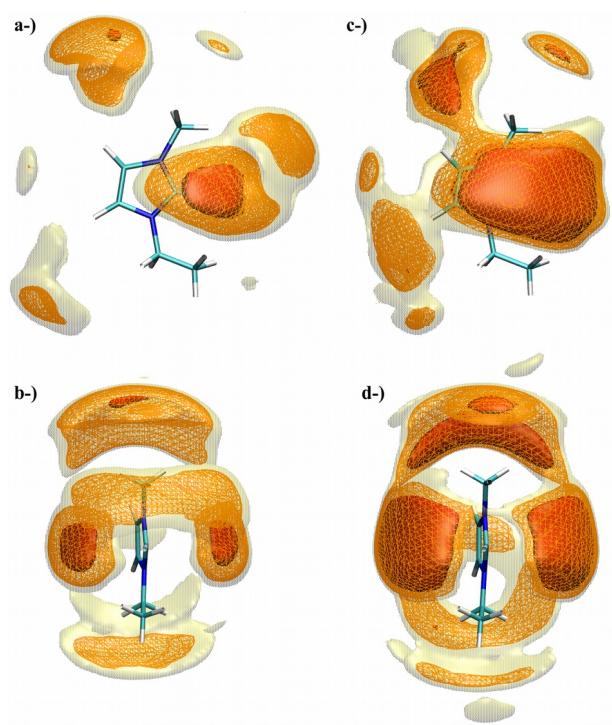

**Figure 9.** Spatial distribution function (sdf) of the NA atom of $[N(CN)_2]^-$ around $[C_2C_1Im]^+$. **a** and **b**: polarizable forcefield; **c** and **d**: non-polarizable forcefield. The top representations (**a** and **c**) are views from above the plane of the imidazolium ring. In the bottom representations (**b** and **d**), the HR atom of imidazolium points toward the reader. Isosurfaces values: 8.5 nm$^{-3}$ (translucid yellow), 9.5 nm$^{-3}$ (orange wireframe) and 12 nm$^{-3}$ (solid red).

The increase of the density from above/below the ring toward to the ring plane in the region close to the HR agrees with the prediction from the ion pair free energy surfaces, as well as the fact that above/below the ring are the most favorable regions for the anion while the region occupied by the ethyl group is the most unfavorable region. As in the previous section, a more direct comparison can be made considering the density of NA only in a sphere of the same radius used for the gas phase ion pair (0.41 nm, Figure 7) and computing the corresponding free energy at each volume element according to Eq. (2). The results are given in Figure 10 where the red-most region corresponds to the smallest free energy obtained and blue regions to the highest ones. As in the case of [BF$_4$]$^-$, the flexibility of the molecules and the interaction involving several ions imply that the surfaces are blurred out and the point to point variations are not so clear as in the gas phase ion pair (compare Figure 10 with representations **a**, **b**, **c** and **d** in Figure 7). However, the essential features remain the same: The lowest free energy region is above/below the imidazolium ring and the free energy close to the plane of the imidazolium ring at the side of the HR atom decreases upon introduction of polarizability in the model (**b** and **d** in Figures 7 and 10), as anticipated by the ion pair surfaces.

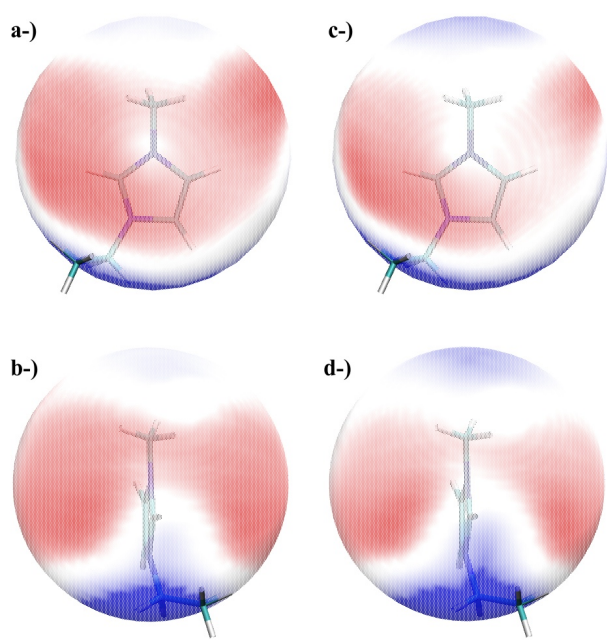

**Figure 10.** Relative free energy surface for the translation of [N(CN)$_2$]$^-$ in a spherical shell of 0.41 nm  around the cation computed in the liquid phase. **a** and **b-)** polarizable force field; **c** and **d-)** non-polarizable force field. The top representations (**a** and **c**) are views from above the plane of the imidazolium ring. In the bottom representations (**b** and **d**), the HR atom of imidazolium points toward the reader. The red regions correspond to the smaller free energy values while the blue corresponds to the largest free energy values of the spherical shell.

A difference between the surfaces for the gas phase ion pair and for the liquid phase that can be directly related to molecular flexibility in [$C_2C_1$Im][DCA] is the more spread and more symmetric high free energy region close to the ethyl group (blue part at the bottom of spheres in Figure 7 and 10 **a** and **c**) in the liquid phase. This is related to the conformational dynamics of the ethyl group, since its rotation changes the region affected by steric hindrance around the cation. In the liquid phase, where one expects a mixture of conformations (see the distributions in Figure S12), there happens an average between the repulsive regions for each conformer that spreads the region with the largest free energy. This effect, therefore, is not directly related to the change from gas phase to the liquid phase, but mostly to the lack of flexibility in the calculations of the gas phase ion pair. In the last case considered, [$C_2C_1$Im][NTf$_2$], the effects of conformations will be even more important since the anion is also flexible and presents two conformers which interact differently with the cation.

## C. 1-ethyl-3-methylimidazolium bis(trifluoromethanesulfonyl)imide, [$C_2C_1$Im][NTf$_2$].

The third IL considered in this work presents an additional challenge since, differently from the [BF$_4$]$^-$ and [N(CN)$_2$]$^-$ that can be considered as rigid species except for small bond and angle deformations, the [NTf$_2$]$^-$ anion is flexible and has two main conformers that have been observed experimentally even in gas phase ion pairs.[44] The most stable exhibits the two CF$_3$ groups at opposite sides of the plane defined by the N and the two S atoms, thus belonging to the C2 symmetry group. Another isomer, with a slightly higher energy, has the two CF$_3$ at the same side of the plane and belongs to the C1 symmetry group. The energy difference between the equilibrium structures of both isomers amounts to 4.1 kJ/mol as obtained by our CAM-B3LYP/Def2-TZVPD calculations, which is close to the 2.6 kJ/mol obtained from the B3LYP/6-311+G(d) calculation by Fujii *et al*.[45] The equilibrium structures of the two conformers are shown in Figure 11, which gives the distribution of the dihedral C-S-S-C computed in the molecular dynamics simulation of the liquid with both of the force fields considered. The global maximum at ± 180° corresponds to the C2 conformer while the local maxima at ± 45° correspond to the C1 conformer.

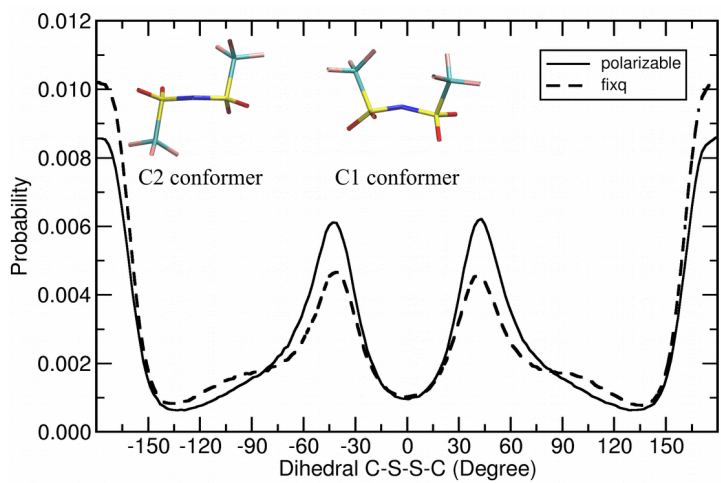

**Figure 11.** Distribution of the dihedral angle defined by the atoms C-S-S-C in [C$_2$C$_1$Im][NTf$_2$]. Equilibrium structures for two conformers of the anion are provided in the figure.

The possibility of two [NTf$_2$]$^-$ conformers with similar probabilities in the liquid phase demands both to be studied in the gas phase ion pair. Thus, separated free energy surfaces were computed for [NTf$_2$]$^-$ in the conformations C1 (Figures S4 for a radius of 0.410 nm and S5 for 0.475 nm) and C2 (Figures S6 for 0.410 nm and S7 for 0.475 nm) using the same method applied so far.

Some variations are remarkable between the free energy surfaces obtained for the two different conformers: For C1, the region at the plane of the ring close to HR presents a higher free energy than the region above/below the ring in both force fields and in DFT calculations (Figures S4 and S5), but for the smaller radius the difference between those regions is overestimated in the non-polarizable force field as it was already observed in the previous cases. The PM7 calculation, on the other hand, seems to underestimate this variation resulting in a more homogeneous surface.

For the C2 conformer (Figures S6 and S7), an inversion happens when the radius of the considered shell around N2 atom increases from 0.410 to 0.475 nm for the DFT calculation (structures **e** and **f**): When the NA atom of [NTf$_2$]$^-$ is forced closer to the N2 of [C$_2$C$_1$Im]$^+$, the lowest free energy region is close to HR at the plane of the ring due to the steric hindrance of the voluminous SO$_2$CF$_3$ groups that are in the opposite sides of the molecules and make difficult the approximation of the N atom above/below the imidazolium ring. The same effect is not present in the surfaces of the C1 conformer since the bulk groups are in the same side of the anion and then a rotation is possible that enables the N atom of getting close to the imidazolium ring from above/below. A similar tendency is found for the polarizable force field and for the PM7 calculations (Figure S6 **a**, **b**, **g**, **h**), being actually overestimated by PM7 in comparison to DFT. For the non-polarizable force field, the region close to the plane of the ring and at the side of HR remains less favorable even though this region has smaller steric hindrance to the approximation (Figure S6 **c** and **d**). For the larger radius (Figure S7), on the other hand, the most favorable region is again above/below the ring and the four methods agree qualitatively well. Notice that the surfaces are noisier for this ion pair due to the complex structure of the anion and the lower symmetry in comparison to the previous ones, so that a larger number of rotations and translations would be needed in order to obtain smooth surfaces for the [NTf$_2$]$^-$ anion.

Even though the surfaces for each [NTf$_2$]$^-$ conformer separately give interesting information, a comparison with liquid phase demands a surface that combines the results from both the conformers. Thus, new surfaces were computed from the data of Figures S4 to S7 according to

$$\Delta F_{comb}(t) = -k_B T \ln\left[\exp\left(-\frac{\Delta F_{C2}(t)}{k_B T}\right) + \exp\left(-\frac{\Delta F_{C1}(t) + \Delta E_{C1-C2}}{k_B T}\right)\right] - F_{comb}(t')$$

(Equation 4)

where $\Delta F_{C2}(t)$ and $\Delta F_{C1}(t)$ are the relative free energies for C2 and C1 calculated in the translational point $t$ separately according to Eq. (1), and $\Delta E_{C1-C2}$ is the energy difference between the isolated conformers. The difference $\Delta E_{C1-C2}$ must be included because in the calculation of the relative free energies by Eq. (1) the smallest free energy is subtracted from every point, hence the energy changes arising from the inner structure of the ions are ruled out in the individual surfaces. The last term in Eq. (3), $F_{comb}(t')$, has the same effect as $F(t')$ in Eq. (1): The smallest free energy found is subtracted from every point in order to set the smallest free energy as the reference zero in the respective surfaces. Thus, Eq. (4) is a convenient way to work out with the already calculated free energy values.

The surfaces including the results from both $[NTf_2]^-$ conformers are given in Figure S8 for the radius of 0.410 nm and in Figure 12 for 0.475 nm. Overall, the combined surfaces exhibit an intermediate pattern between the ones computed for each conformer separately, but it is not a simple average as follows from Eq. (3): If in a given point $t$ the value of $\Delta F_{C2}(t)$ is much smaller than the one of $\Delta F_{C1}(t) + \Delta E_{C1-C2}$, then $\Delta F_{comb}(t)$ will be close to $\Delta F_{C2}(t)$ while, if the opposite is true, $\Delta F_{comb}(t)$ will be close to $\Delta F_{C1}(t) + \Delta E_{C1-C2}$. For both the radius and in every method employed, the most favorable region is below/above the plane of the imidazolium ring, but both the force fields and the semi-empirical calculations underestimate it and results in more homogeneous surfaces than the DFT calculation for the radius of 0.475 nm (Figure 12). For the smaller radius (Figure S8), the polarizable model agrees well with DFT (Figure S8 **a**, **b**, **e**, and **f**), but the non-polarizable one overestimates the free energy differences (Figure S8 **c** and **d**), resulting in a larger relative value close to the plane of the ring in the side of HR atom, while the PM7 method underestimates it (Figure S8 **g** and **h**). The lowest energy structures for C1 and C2 conformers are given in Figures S16 and S17 and have similar features as the MP2 optimizations performed by Stearns *et al*,[43] with the anion above/below the ring with one of the oxygen atoms close to the HR of the cation.

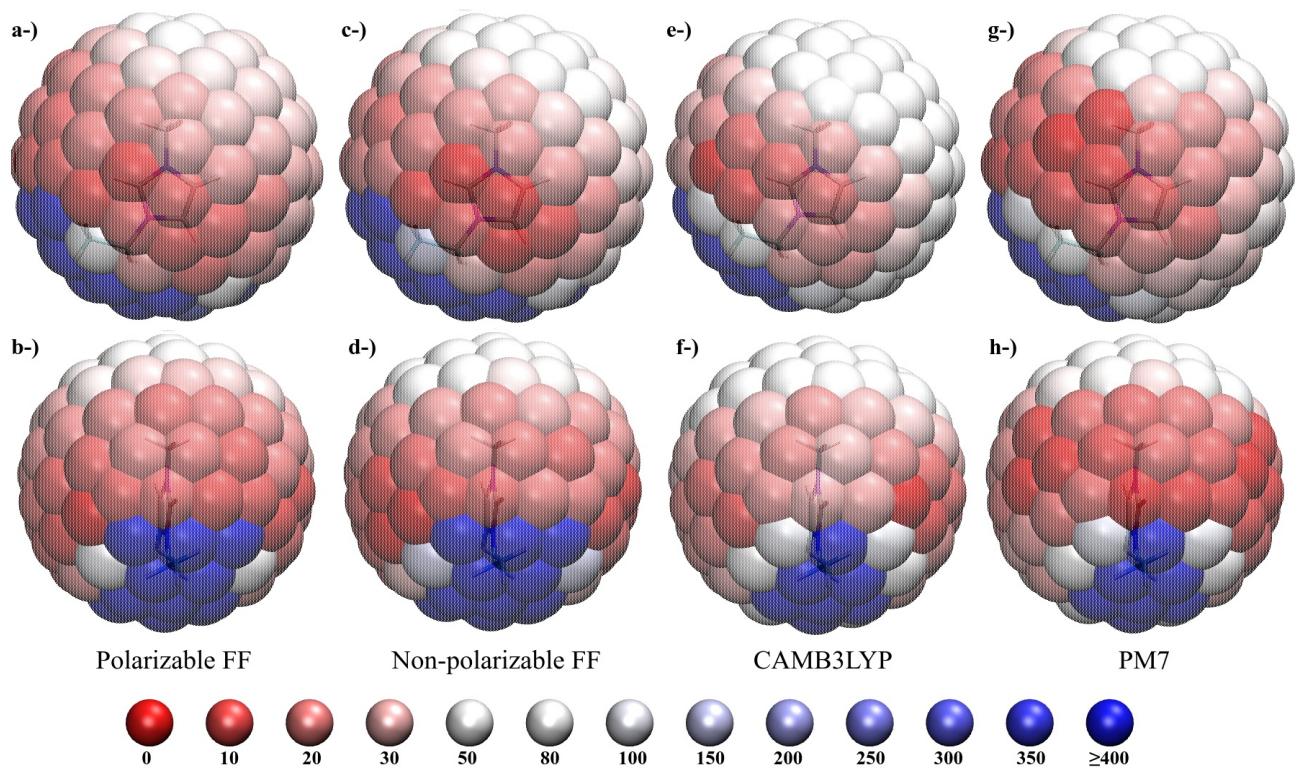

**Figure 12.** Free energy surfaces for the [$C_2C_1Im$][$NTf_2$] ion pair with the distance N2-NA of 0.475 nm. The colors correspond to different free energy values according to the legend in the bottom (kJ/ mol). **a** and **b**-) Polarizable forcefield; **c** and **d**-) non-polarizable forcefield; **e** and **f**-) DFT calculation; **g** and **h**-) semi-empirical calculation. The top representations are views from above the plane of the imidazolium ring. In the bottom representations, the HR atom of imidazolium points toward the reader. See Supporting Information for other distances.

The radial distribution functions for the liquid phase, as in the case of [$C_4C_1Im$][$BF_4$], display only small variations between the polarizable and non-polarizable models in the first coordination shell (see Figure 13). The larger difference is observed for the HR atom (red curves), which presents a larger density of NA atom at the distance of 0.25 nm when the non-polarizable force field is used. A similar result was discussed for the [$C_2C_1Im$][$N(CN)_2$] in Figure 8, but it happens for a different reason in [$C_2C_1Im$][$NTf_2$]: the radial distribution functions of other atoms of [NTf$_2$]$^-$ around the HR (Figure S13) show no significant increase in the first peak when polarizability is introduced, so this cannot be attributed to a change of preference for one atom of anion over another as it was in the case of the [N(CN)$_2$]$^-$ anion. Instead, here the change in the $g(r)$ can be understood by the spatial distribution function of NA around the cation (Figure 14) or by the subtraction of the sdfs obtained with the two force fields (Figure S20). In contrast to the previous systems, there is only a small increase of density in the plane of the ring and close to the HR atom when polarizability is introduced in [C$_2$C$_1$Im][NTf$_2$]. However, there is a more pronounced reduction of the density close to the HR but out of the plane in the polarizable force field (this is more clearly seen in the subtraction shown in Figure S20, where the red isosurfaces up and below the HR demonstrate a smaller density of NA in the polarizable model). It should be noted that the same two effects, that is, increase of NA density in the plane of the ring and decrease above/below the ring when including polarization, is observed for every IL considered here. However, while for [C$_4$C$_1$Im][BF$_4$] (Figure S18) the two effects have similar weights, resulting in overall no change in $g(r)$ (Figure 4), for [C$_2$C$_1$Im][NTf$_2$] the increase at the plane is almost negligible and the second effect dominates, resulting in a decrease of the $g(r)$ of NA around HR for short distances. For [C$_2$C$_1$Im][N(CN)$_2$], as discussed before, there is also an effect over the relative orientation of the anion, which results in larger changes over the computed $g(r)$ (Figure 8).

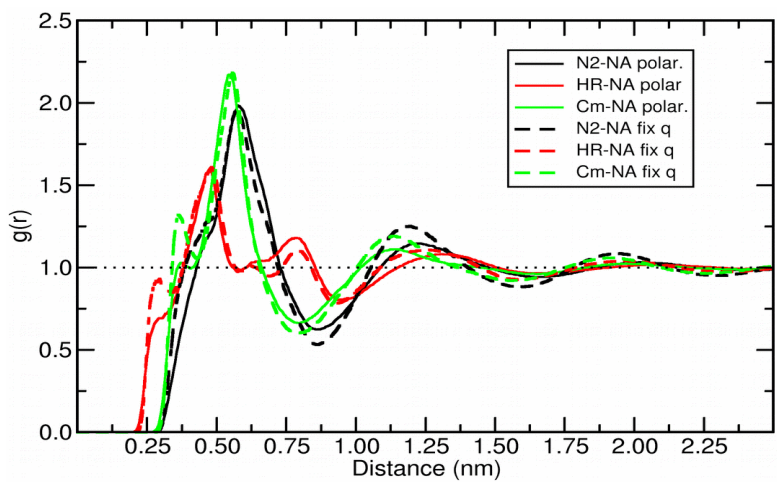

**Figure 13.** Radial distribution function, $g(r)$, for the NA atom of $[NTf_2]^-$ around selected atoms of $[C_2C_1Im]^+$ calculated from liquid simulations with polarizable (solid lines) and non-polarizable (dashed lines) force fields.

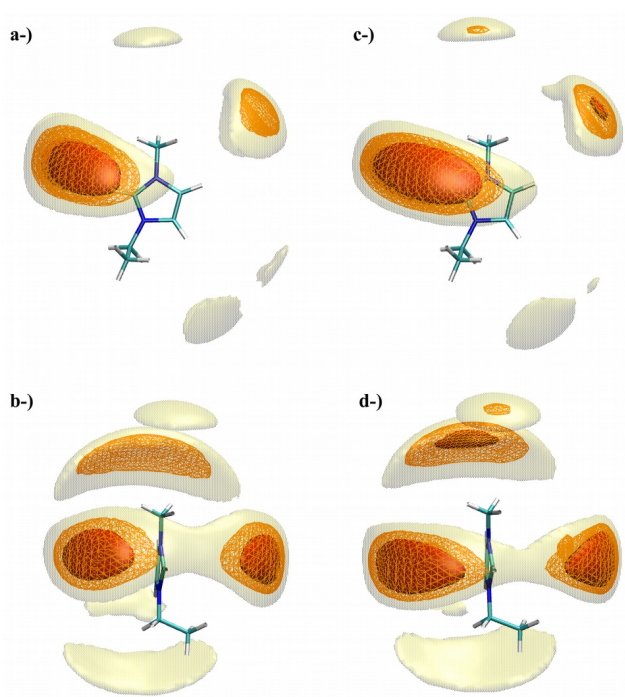

**Figure 14.** Spatial distribution function (sdf) of the NA atom of $[NTf_2]^-$ around $[C_2C_1Im]^+$. **a** and **b**: polarizable force field; **c** and **d**: non-polarizable force field. The top representations (**a** and **c**) are views from above the plane of the imidazolium ring. In the bottom representations (**b** and **d**), the HR atom of imidazolium points toward the reader. Isosurfaces values: 6 $nm^{-3}$ (translucid yellow), 8 $nm^{-3}$ (orange wireframe) and 10 $nm^{-3}$ (solid red).

The change of [NTf$_2$]$^-$ density close to the plane of the imidazolium ring in the side of HR upon inclusion of polarization is small in comparison to the previous examples. This finding is in line with the combined free energies surfaces for the ion pair, that display almost no free energy difference between the two force fields in this region (Figure 12 **a**, **b**, **c**, **d**). Except for the case of small radius (Figure S8), for which the overall contribution should be smaller according to the small $g(r)$ in this region (Figure 13), the surfaces for the polarizable and non-polarizable force field are very similar for this ion pair. Accordingly, Figure 15 shows that the free energy surfaces for the liquid phase calculated by using Eq. (3) do not indicate significant variations in the first shell structure between the two force fields.

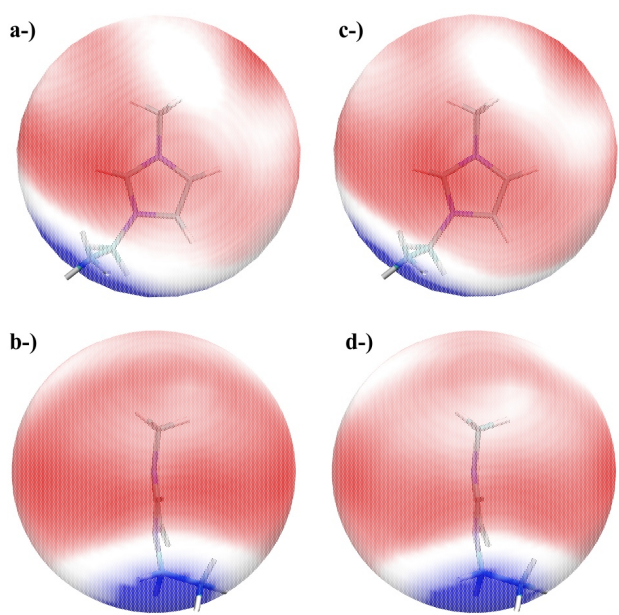

**Figure 15.** Relative free energy surface for the translation of [NTf$_2$]$^-$ in a spherical shell of 0.41 nm around [C$_2$C$_1$Im]$^+$ calculated in the liquid phase. **a** and **b-)** polarizable force field; **c** and **d-)** non-polarizable force field. The top representations (**a** and **c**) are views from above the plane of the imidazolium ring. In the bottom representations (**b** and **d**), the HR atom of imidazolium points toward the reader. The red regions correspond to the smaller free energy values while the blue corresponds to the largest free energy values of the spherical shell.

It should be noted, however, that the overall match between the free energy surfaces computed for gas phase ion pair (Figure 12 **a**, **b**, **c**, **d**) and the corresponding liquid phase surfaces (Figure 15) for $[C_2C_1Im][NTf_2]$ is not as good as it was for the other liquids considered. Two main effects can explain the worse agreement here. First, the flexibility of $[NTf_2]^-$. Even though the equilibrium geometry of the two isomers were considered for the gas phase ion pair, several intermediate structures should be considered as well since they also have relatively large probabilities to be found in the condensed phase (Figure 11). Second, larger size and smaller symmetry of $[NTf_2]^-$ result in surfaces for the gas phase ion pair noisier than observed in the previous two cases, so that more rotations and precessions movements would be necessary to improve the quality of the gas phase surfaces.

All in all, the use of free energy surfaces calculated for gas phase ion pairs by the presented methodology and the structure of the first coordination shell of the anion around the cation are very similar for rigid and relatively small anions such as $[BF_4]^-$ and $[N(CN)_2]^-$, enabling the comparison between structures predicted by classical force fields and the ones that would be obtained if a high level electronic structure  calculation could be used for the simulation of the liquid. In the case of flexible and large anions like $[NTf_2]^-$ some extra considerations are needed, in particular the consideration of conformational changes for the free energy calculations of the ion pair.

## IV. CONCLUSIONS

The calculation of free energy surfaces for the gas phase ion pair evidences the differences between the several theory levels employed. The non-polarizable force field underestimates the interaction when the anion is at the plane of imidazolium ring in relation to the structures where the anion is located above/below the ring for every pair considered. This effect is corrected at least partially when polarizability is introduced using Drude induced dipoles. Despite some remaining differences between results from the polarizable model and the reference DFT calculations, the overall performance of the polarizable force field is as good as the one of the semi-empirical method PM7 and even better in some cases.

Most of the structural trends observed from the free energy surfaces for the gas phase ion pair is manifest in the structure of the first coordination shell in the liquid phase, even though the free energy surfaces calculated for the liquid get blurrier due to molecular flexibility and the many-body effects in the condensed phase. The agreement of free energy surfaces for the gas and liquid phases was better for the rigid and more symmetric anions $[BF_4]^-$ and $[(N(CN)_2]^-$ than for the flexible and low-symmetry $[NTf_2]^-$, for which intermediary conformations besides the two equilibrium geometries should be considered to improve the results in the gas phase calculations.

On one hand, the agreement of the trends in the structure of the first shell between the gas and liquid phases for the ionic liquids under consideration showed that the proposed methodology can be used to predict the short range structure of ionic liquids employing high-level theory methods that are prohibitive to simulations of condensed phases with the current computational power. On the other hand, the description obtained with a high-level calculation can be compared with the one obtained from low-level or force fields methods in order to select the one that gives the best description of the short-range structure or even to indicate possible improvements in the parametrization of available force fields.

**SUPPORTING INFORMATION**

Figures with additional free energy surfaces, radial and angle distribution functions, the lowest energy structures obtained by DFT calculations, and differences in spatial distribution functions between polarizable and non-polarizable models.

**ACKNOWLEDGMENTS**

The Brazilian authors are indebted to FAPESP (Grants 2019/04785-9, 2017/12063-8, 2016/21070-5) and CNPq (Grant 301553/2017-3) for financial support and the "Laboratório Nacional de Computação Científica (LNCC/MCTI, Brazil)" for the use of the supercomputer SDumont (https://sdumont.lncc.br). The French authors thank the IDEX Lyon Fellowship (ANR-16-IDEX-005) and the use of the computer clusters of the Pôle Scientifique de Modélisation Numérique (PSMN) at ENS de Lyon. We also thanks Dr. Felippe M. Colombari from "Universidade Federal de São Carlos", São Carlos, SP, Brazil, to provide a copy of the software THEMIS (manuscript under submission).

**REFERENCES**

1. E. J. Maginn "Molecular simulation of ionic liquids: current status and future opportunities". *J. Phys.: Condens. Matter* **21**, 373101 (2009).

2. J. Canongia Lopes, J. Deschamps and A. A. H. Pádua "Modeling Ionic Liquids Using a Systematic All-Atom Force Field" *J. Phys. Chem. B* **108**, 2038–2047 (2008).

3. J. N. Canongia Lopes and A. Pádua "CL&P: A generic and systematic force field for ionic liquids modeling" *Theor. Chem. Acc.* **131**, 1129 (2012).

4. S. V Sambasivarao and O. Acevedo "Development of OPLS-AA Force Field Parameters for 68 Unique Ionic Liquids" *J. Chem. Theory Comput.* **5**, 1038–1050 (2009).

5. Y. Zhang; E. J. Maginn "Direct Correlation Between Ionic Liquid Transport Properties and Ion Pair Lifetimes: A Molecular Dynamics Study". *J. Phys. Chem. Lett.* **6**, 700−705 (2015).

6. F. Khabaz, Y. Zhang, L. Xue, E. L. Quitevis, E. J. Maginn and R. Khare "Temperature Dependence of Volumetric and Dynamic Properties of Imidazolium-Based Ionic Liquids" *J. Phys. Chem. B* **122**, 2414−2424 (2018).

7. Y. Zhang; E. J. Maginn "The effect of C2 substitution on melting point and liquid phase dynamics of imidazolium based-ionic liquids: insights from molecular dynamics simulations". *Phys. Chem. Chem. Phys.* **14**, 12157–12164 (2012).

8. J. G. McDaniel and A. Yethiraj "Influence of Electronic Polarization on the Structure of Ionic Liquids" *J. Phys. Chem. Lett.* **9**, 4765−4770 (2018).

9. C. Schröder, T. Sonnleitner, R. Buchner and O. Steinhauser "The influence of polarizability on the dielectric spectrum of the ionic liquid1-ethyl-3-methylimidazolium triflate" *Phys Chem Chem Phys* **13**, 12240 (2011).

10. V. H. Paschoal and M. C. C. Ribeiro. "Molecular dynamics simulations of high-frequency sound modes in ionic liquids". *J. Mol. Liq.* **202**, 252-256 (2015).

11. A. A. Veldhorst and M. C. C. Ribeiro. "Mechanical heterogeneity in ionic liquids". *J. Chem. Phys.* **148**, 193803 (2018).

12. D. Bedrov, J.-P. Piquemal, O. Borodin, A. D. MacKerell, B. Roux and C. Schröder "Molecular Dynamics Simulations of Ionic Liquids and Electrolytes Using Polarizable Force Fields". *Chem. Rev.*, **119**, 7940–7995 (2019).

13. B. L Bhargava and S. J. Balasubramanian "Refined potential model for atomistic simulations of ionic liquid [bmim][PF6]". *J. Chem. Phys.* **127**, 114510 (2007).

14. V. Chaban "Polarizability versus mobility: atomistic force field for ionic liquids". *Phys. Chem. Chem. Phys.* **13**, 16055–16062 (2011).

15. Y. Zhang and E. J. Maginn "A Simple AIMD Approach to Derive Atomic Charges for Condensed Phase Simulation of Ionic Liquids". *J. Phys. Chem. B* **116**, 10036–10048 (2012).

16. K. Goloviznina, J. N. C. Lopes, M. C. Gomes and A. Padua, "A Transferable, Polarisable Force Field for Ionic Liquids", *J. Chem. Theory Comput.*. DOI:10.26434/chemrxiv.8845526.v2 (2019).

17. C. Schröder "Comparing reduced partial charge models with polarizable simulations of ionic liquids" *Phys. Chem. Chem. Phys.* **14**, 3089–3102 (2012).

18. O. Borodin "Polarizable Force Field Development and Molecular Dynamics Simulations of Ionic Liquids" J. *Phys. Chem. B* **113**, 11463–11478 (2009).

19. J. G. McDaniel, E. Choi, C. Y. Son, J. R. Schmidt and A. Yethiraj "Ab Initio Force Fields for Imidazolium-Based Ionic Liquids" *J. Phys. Chem. B* **120**, 7024–7036 (2016).

20. W. L. Jorgensen, D. S. Maxwell and J. Tirado-Rives "Development and Testing of the OPLS All-Atom Force Field on Conformational Energetics and Properties of Organic Liquids" *J. Am. Chem. Soc.* **118**, 11225–11236 (1996).

21. H. Weber; B. Kirchner "Complex Structural and Dynamical Interplay of Cyano-Based Ionic Liquids." *J. Phys. Chem. B* **120**, 2471-2483 (2016).

22. R. S. Booth; C. J. Annesley; J. W. Young; K. M. Vogelhuber; J. A. Boatz and J. A. Stearns . "The Nature of Ionic Liquids in the Gas Phase". *J. Phys. Chem. A* **111**, 6176-6182 (2007).

23. M. Sun; L. Xu, A. Qu, P. Zhao, T. Hao, W. Ma, C. Hao, X. Wen, F. M. Colombari, A. F. de Moura, N. A. Kotov, C. Xu and H. Kuang. "Site-selective photoinduced cleavage and profiling of DNA by chiral semiconductor nanoparticles". *Nature Chemistry* **10**, 821–830 (2018).

24. T. Yanai, D. P. Tew, N. C. Handy "A new hybrid exchange–correlation functional using the Coulomb-attenuating method (CAM-B3LYP)" *Chem. Phys. Lett.* **393**, 51–57 (2004).

25. F. Weigend and R. Ahlrichs "Balanced basis sets of split valence, triple zeta valence and quadruple zeta valence quality for H to Rn: Design and assessment of accuracy". *Phys. Chem. Chem. Phys.* **7**, 3297 (2005).

26. S. Grimme, S. Ehrlich, L. Goerigk "Effect of the damping function in dispersion corrected density functional theory." *J Comput Chem.* **32**, 1456–1465 (2011).

27. F. Neese The ORCA program system, Wiley Interdiscip. Rev.: Comput. Mol. Sci. 2, 73-78 (2012).

28. G. Lamoureux and B. Roux, "Modeling induced polarization with classical Drude oscillators: Theory and molecular dynamics simulation algorithm" *J. Chem. Phys.* **119**, 3025–3039 (2003).

29. S. Plimpton "Fast Parallel Algorithms for Short-Range Molecular Dynamics" *J. Comput. Phys.* **117**, 1–19 (1995).

30. MOPAC2016, James J. P. Stewart, Stewart Computational Chemistry, Colorado Springs, CO, USA, HTTP://OpenMOPAC.net (2016).

31. A. A. H. Padua, fftool, http://github.com/agiliopadua/fftool.

32. L. Martínez, R. Andrade, E. G. Birgin and J. M. Martínez "PACKMOL: a package for building initial configurations for molecular dynamics simulations." *J. Comput. Chem.* **30**, 2157–2164 (2009).

33. J. N. Canongia Lopes and A. A. H. Pádua "Molecular Force Field for Ionic Liquids Composed of Triflate or Bistriflylimide Anions". *J. Phys. Chem. B* **108**, 16893–16898 (2004).

34. A. A. H. Padua, ilff, http://github.com/agiliopadua/ilff.

35. A. S. L. Gouveia, C. E. S. Bernardes, L. C. Tomé, E. I. Lozinskaya, Y. S. Vygodskii, A. S. Shaplov, J. N. C. Lopes and I. M. Marrucho "Ionic liquids with anions based on fluorosulfonyl derivatives: from asymmetrical substitutions to a consistent force field model". *Phys. Chem. Chem. Phys.* **19**, 29617–29624 (2017).

36. E. Heid, A. Szabadi and C. Schröder "Quantum mechanical determination of atomic polarizabilities of ionic liquids". *Phys. Chem. Chem. Phys.* **20**, 10992–10996 (2018).

37. S. Y. Noskov, G. Lamoureux and B. Roux "Molecular Dynamics Study of Hydration in Ethanol−Water Mixtures Using a Polarizable Force Field" *J. Phys. Chem. B* **109**, 6705–6713 (2005).

38. A. Dequidt, J. Devémy and A. A. H. Padua "Thermalized Drude Oscillators with the LAMMPS Molecular Dynamics Simulator" *J. Chem. Inf. Model.* **56**, 260–268 (2015).

39. A. A. H. Pádua "Resolving dispersion and induction components for polarisable molecular simulations of ionic liquids" *J. Chem. Phys.* **146**, 204501 (2017).

40. T. M. Parker, L. A. Burns, R. M. Parrish, A. G. Ryno and C. D. Sherrill "Levels of symmetry adapted perturbation theory (SAPT). I. Efficiency and performance for interaction energies." *J. Chem. Phys.* **140**, 94106 (2014).

41. M. Brehm and B. Kirchner "TRAVIS - A free Analyzer and Visualizer for Monte Carlo and Molecular Dynamics Trajectories". *J. Chem. Inf. Model.* **51** (8), 2007-2023 (2011).

42. W. Humphrey, A. Dalke and K. Schulten, "VMD – Visual Molecular Dynamics". *J. Molec. Graphics* **14.1**, 33-38 (1996).

43. K. Dong; S. Zhang; D. Wang and X. Yao. "Hydrogen bonds in imidazolium ionic liquids". *J. Phys. Chem. A* **110**, 9775-9782 (2006).

44. R. S. Booth, C. J. Annesley, J. W. Young, K. M. Vogelhuber, J. A. Boatz and J. A. Stearns "Identification of multiple conformers of the ionic liquid [emim][tf2n] in the gas phase using IR/UV action spectroscopy" *Phys. Chem. Chem. Phys.* **18**, 17037 (2016).

45. K. Fujii; T. Fujimori; T. Takamuku; R. Kanzaki; Y. Umebayashi and S. Ishiguru. "Conformational Equilibrium of Bis(trifluoromethanesulfonyl) Imide Anion of a Room-Temperature Ionic Liquid: Raman Spectroscopic Study and DFT Calculations". *J. Phys. Chem. B Lett.* **110**, 8179-8183 (2006).

**SUPPORTING INFORMATION**

**Ion Pair Free Energy Surface as a Probe of Ionic Liquid Structure**

Kalil Bernardino,[1*] Kateryna Goloviznina,[2]

Agílio A. H. Padua,[2] Margarida Costa Gomes,[2] Mauro C. C. Ribeiro[1]

[1] *Laboratório de Espectroscopia Molecular, Departamento de Química Fundamental,*
*Instituto de Química, Universidade de São Paulo, Av. Prof. Lineu Prestes 748, 05508-000, Brazil*
[2] *Univ Lyon, Ens de Lyon, CNRS UMR 5182, Université Claude Bernard Lyon 1,*
*Laboratoire de Chimie, F69342, Lyon, France*

*email: kalil.bernardino@gmail.com

Contents:

## 1. Addtional Free Energy Surfaces

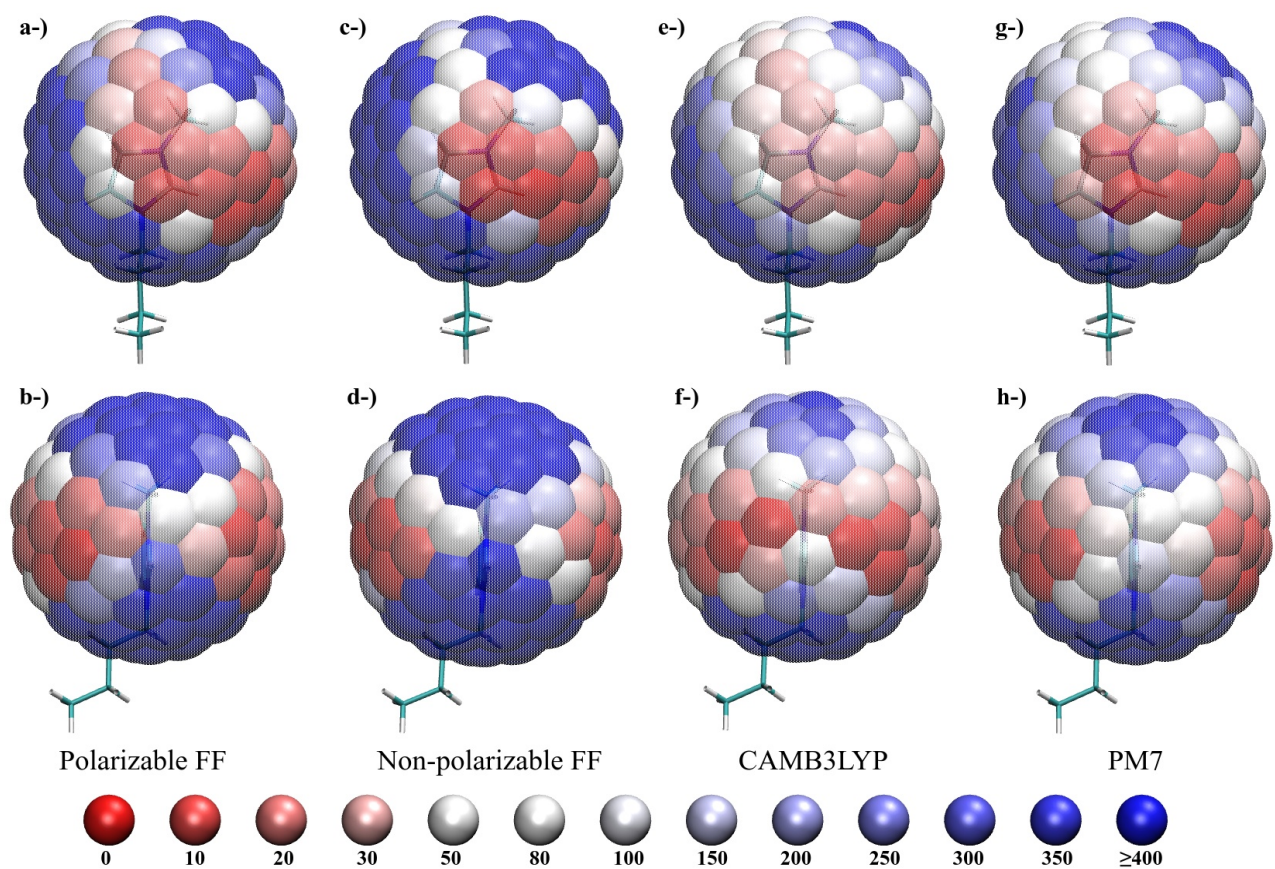

**Figure S1.** Free energy surfaces for the [C₄C₁Im][BF₄] ion pair with the distance N2-B of 0.365 nm. The different colors correspond to different free energy values according to the legend in the bottom (kJ/mol).

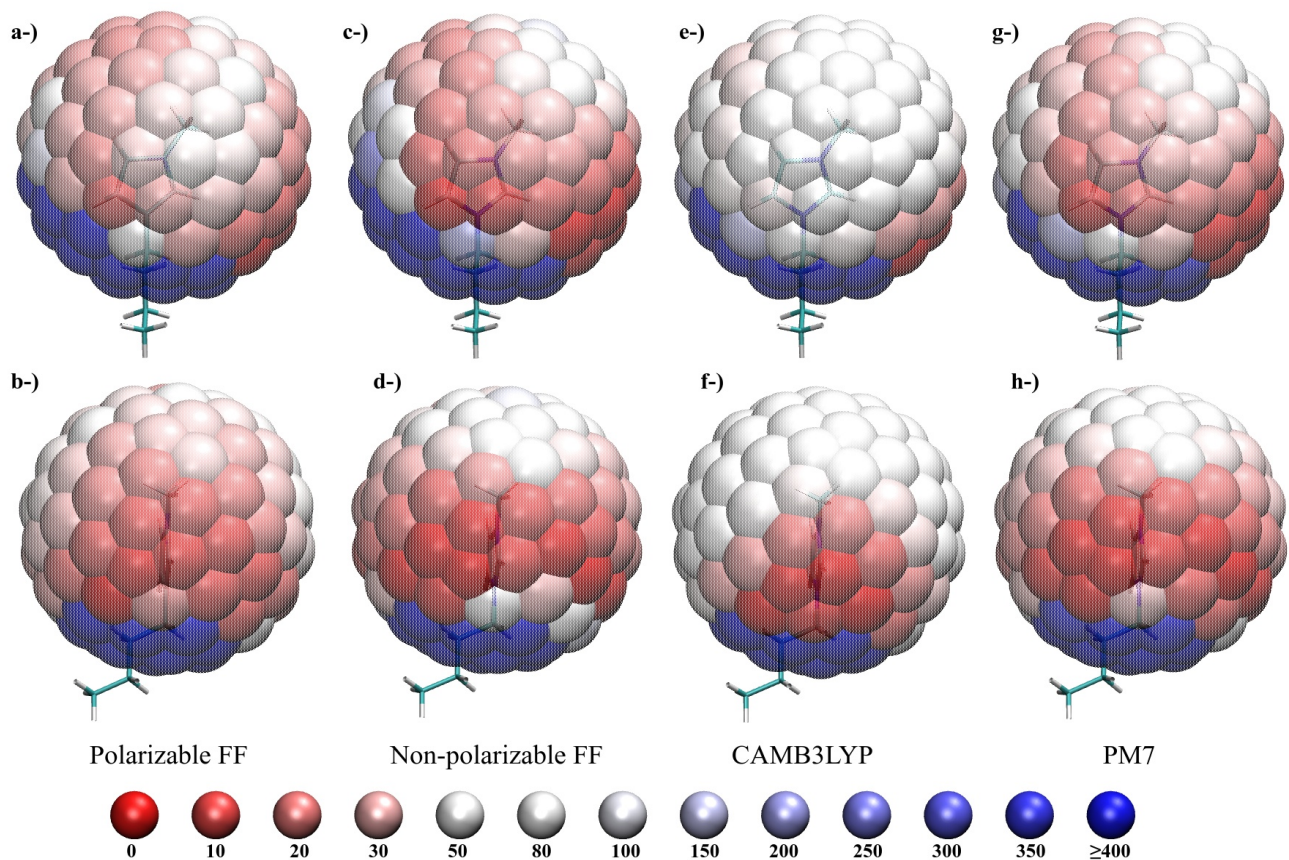

**Figure S2.** Free energy surfaces for the [C₄C₁Im][BF₄] ion pair with the distance N2-B of 0.44 nm. The different colors correspond to different free energy values according to the legend in the bottom (kJ/mol).

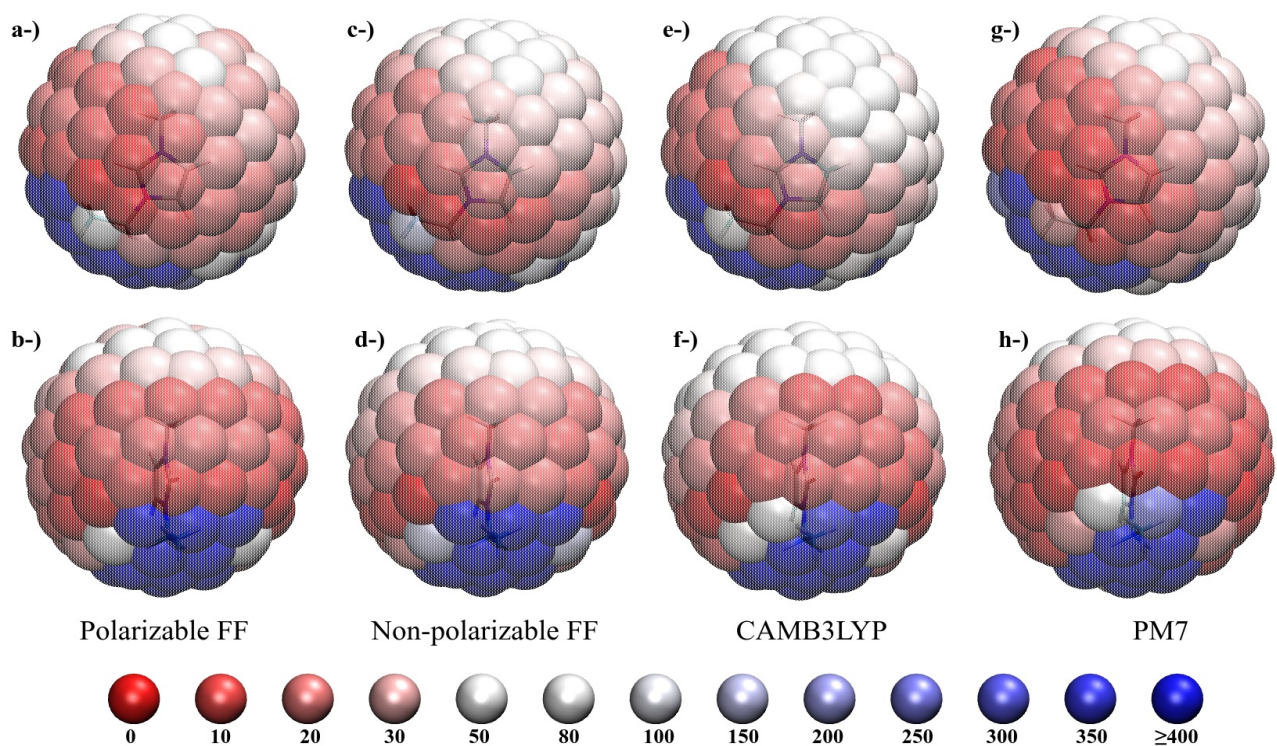

**Figure S3.** Free energy surfaces for the [C$_2$C$_1$Im][N(CN)$_2$] ion pair with the distance N2-N of 0.47 nm. The different colors correspond to different free energy values according to the legend in the bottom (kJ/mol).

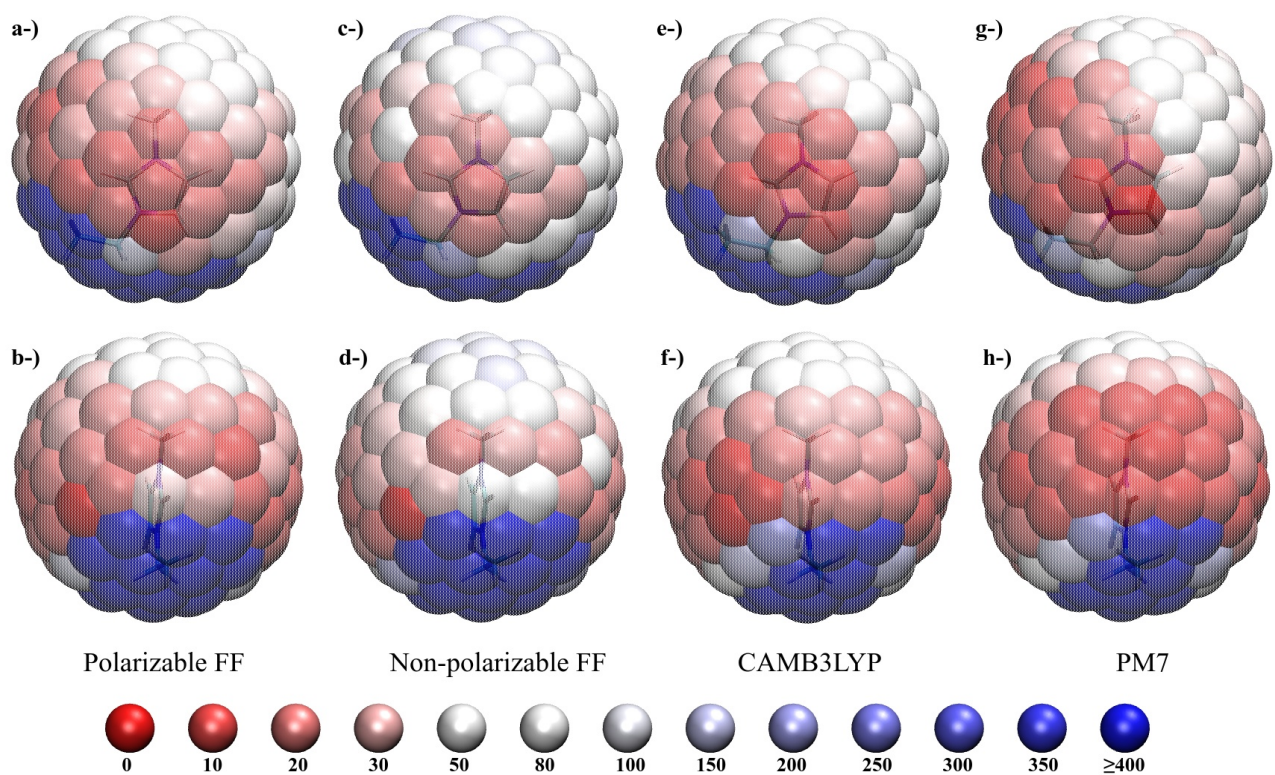

**Figure S4.** Free energy surfaces for the [C₂C₁Im][NTf₂] ion pair with the distance N2-N of 0.41 nm and the anion in the conformation C1. The different colors correspond to different free energy values according to the legend in the bottom (in kJ/mol).

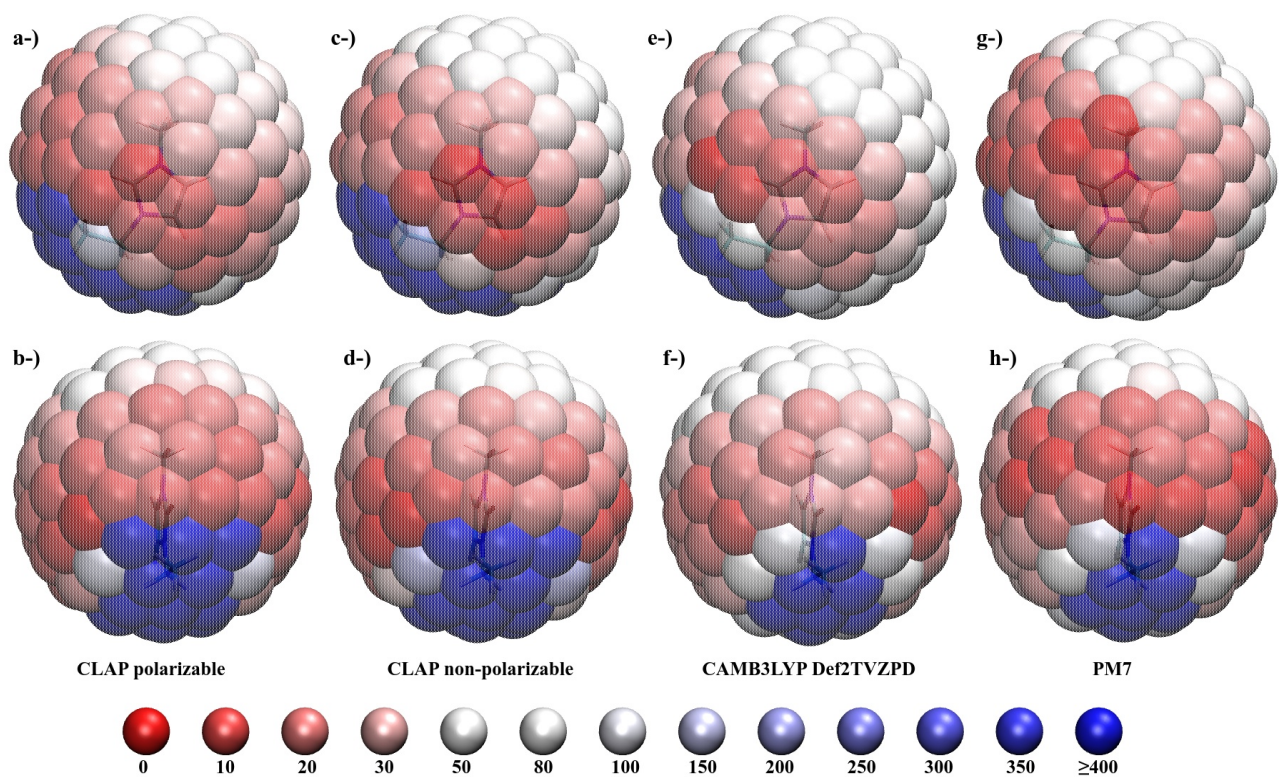

**Figure S5.** Free energy surfaces for the [C$_2$C$_1$Im][NTf$_2$] ion pair with the distance N2-N of 0.475 nm and the anion in the conformation C1. The different colors correspond to different free energy values according to the legend in the bottom (kJ/mol).

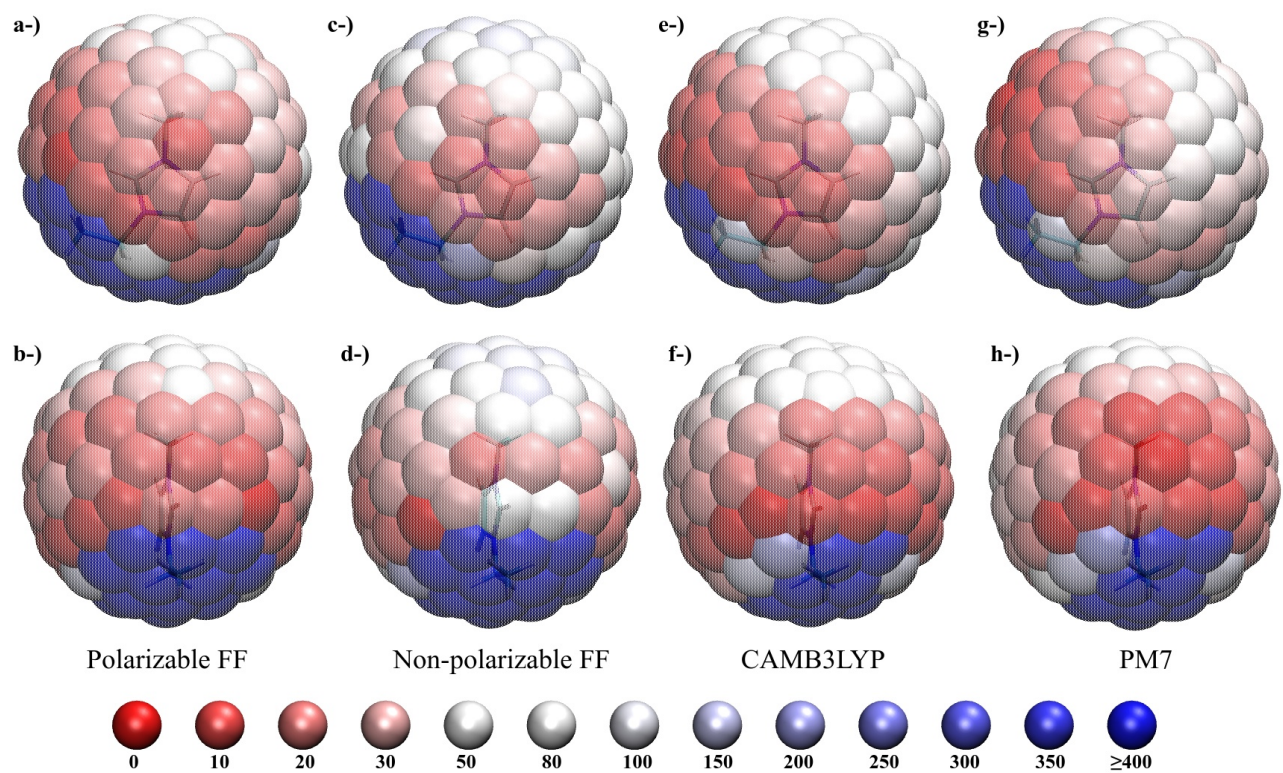

**Figure S6.** Free energy surfaces for the [C₂C₁Im][NTf₂] ion pair with the distance N2-N of 0.41 nm and the anion in the conformation C2. The different colors correspond to different free energy values according to the legend in the bottom (kJ/mol).

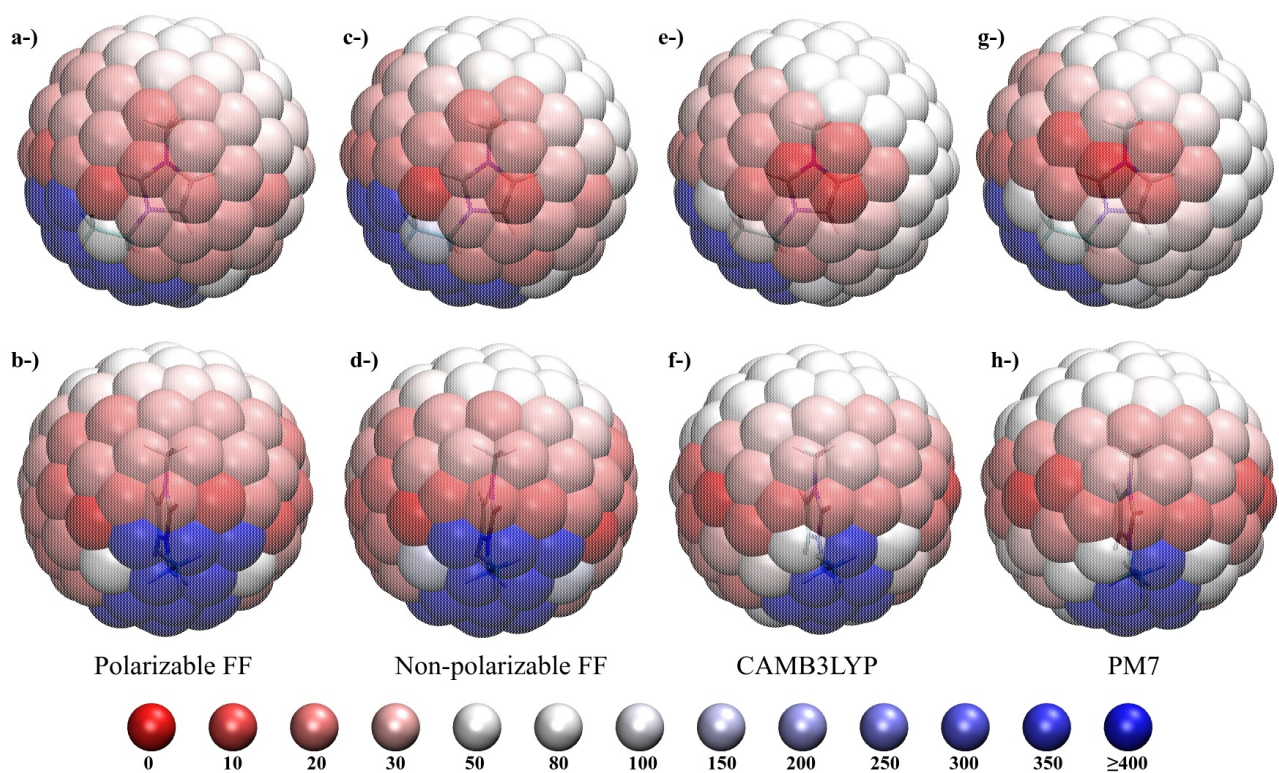

**Figure S7.** Free energy surfaces for the [C₂C₁Im][NTf₂] ion pair with the distance N2-N of 0.475 nm and the anion in the conformation C2. The different colors correspond to different free energy values according to the legend in the bottom (kJ/mol).

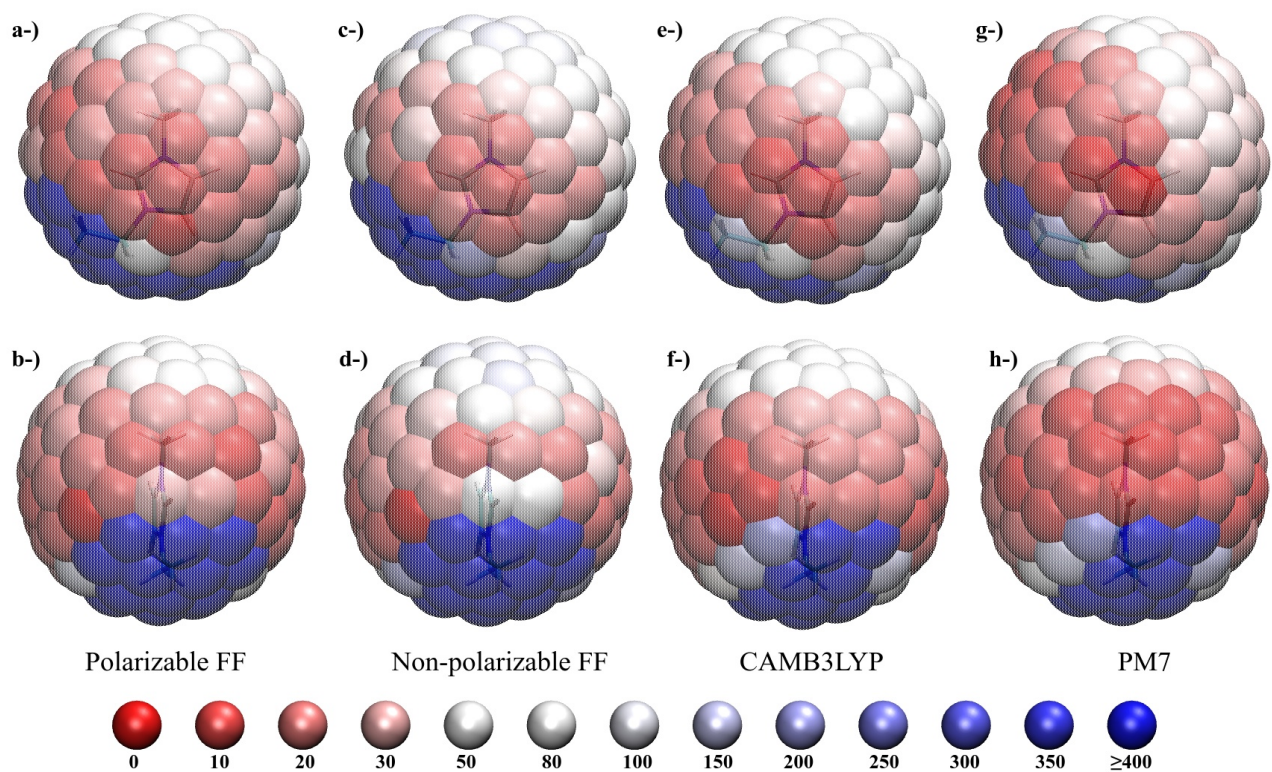

**Figure S8.** Free energy surfaces for the [C$_2$C$_1$Im][NTf$_2$] ion pair with the distance N2-N of 0.41 nm with combined results for the anion in the two conformations. The different colors correspond to different free energy values according to the legend in the bottom (kJ/mol).

## 2. Addtional Distribution Functions

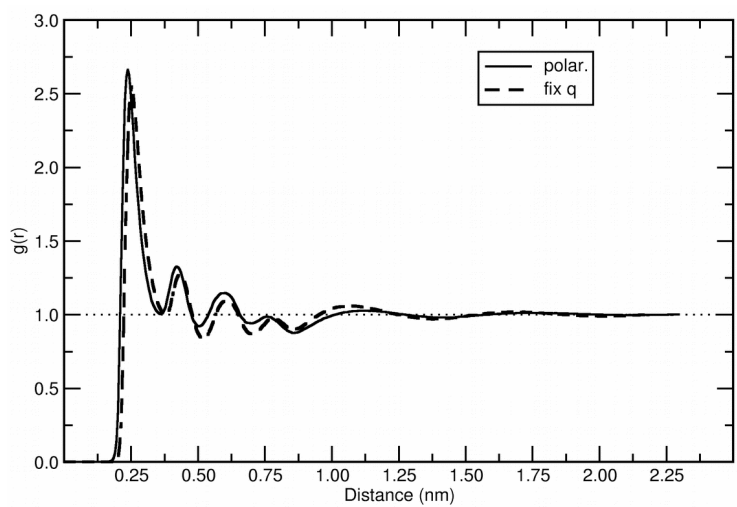

**Figure S9.** Radial distribution function of fluorine atom of [BF$_4$]$^-$ around the HR atom in the [C$_4$C$_1$Im][BF$_4$] liquid.

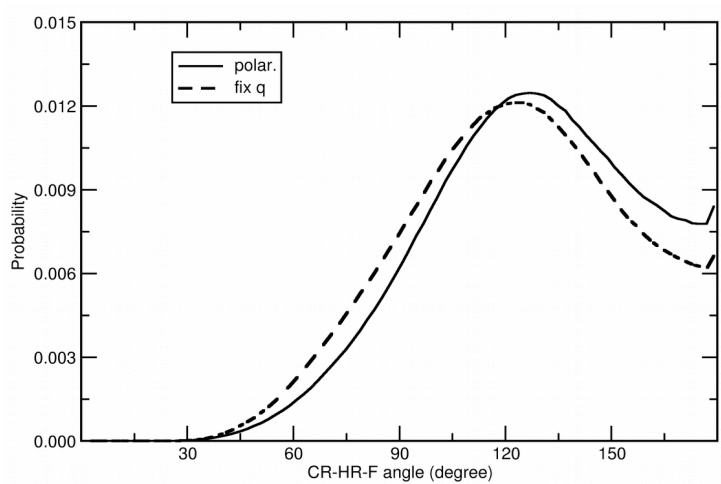

**Figure S10.** CR-HR-F angle distribution obtained in the simulations with the polarizable (solid lines) and non-polarizable (dashed lines) models for the fluorine atoms within the first coordination shell of HR (0.369 nm for polarizable and 0.376 for non-polarizable force fields).

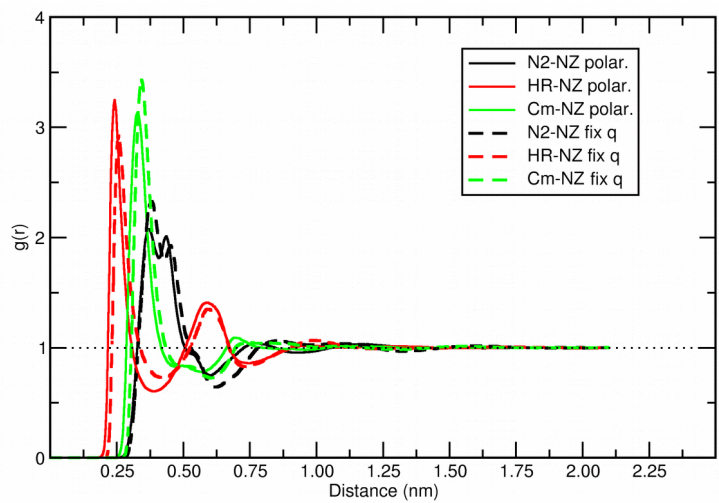

**Figure S11.** Radial distribution function of terminal nitrogen atoms of $[N(CN)_2]^-$ around the selected atoms in the $[C_2C_1Im][N(CN)_2]$ liquid.

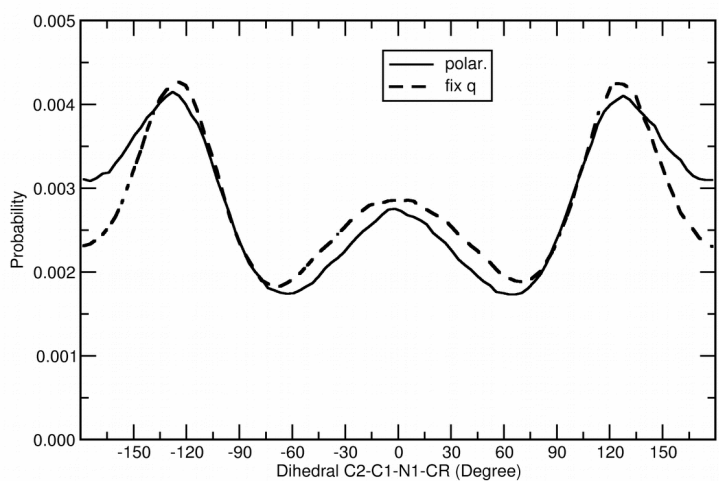

**Figure S12.** Distribution of the dihedral angle C2-C1-N1-CR for the cation in $[C_2C_1Im][N(CN)_2]$ liquid.

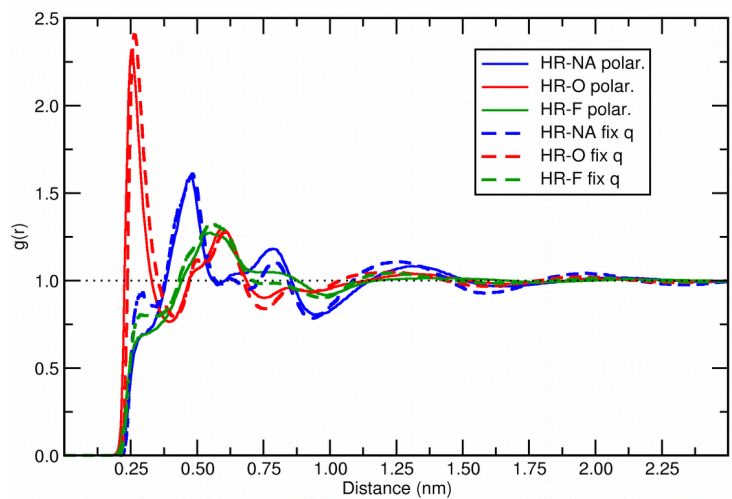

**Figure S13.** Radial distribution function of N, O and F atoms of the anion around the hydrogen HR of the cation in the liquid [C$_2$C$_1$Im][NTf$_2$] simulations with polarizable (solid lines) and non-polarizable (dashed lines) force fields.

**3. Lowest Energy Structures Found in Ion Pair Calculations**

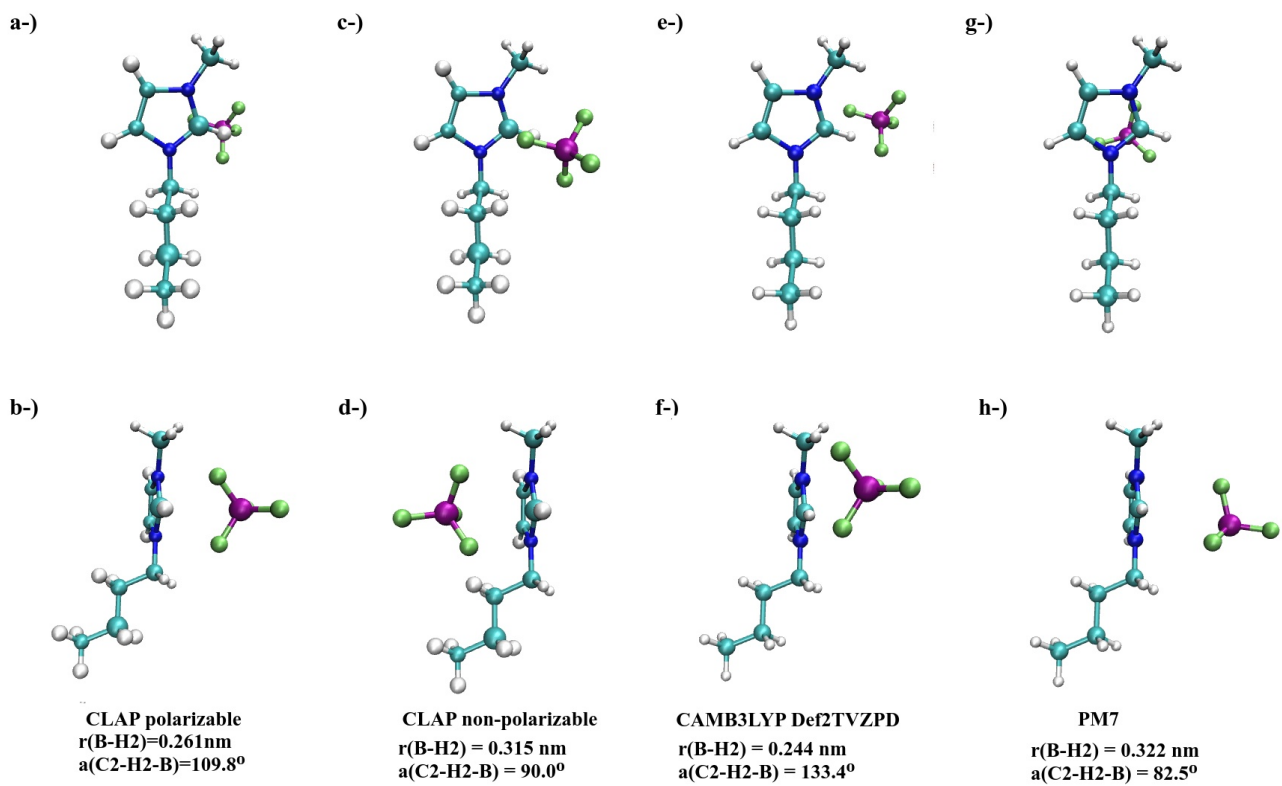

| | | | |
|---|---|---|---|
| **CLAP polarizable** | **CLAP non-polarizable** | **CAMB3LYP Def2TVZPD** | **PM7** |
| **r(B-H2)=0.261nm** | **r(B-H2) = 0.315 nm** | **r(B-H2) = 0.244 nm** | **r(B-H2) = 0.322 nm** |
| **a(C2-H2-B)=109.8º** | **a(C2-H2-B) = 90.0º** | **a(C2-H2-B) = 133.4º** | **a(C2-H2-B) = 82.5º** |

**Figure S14.** Lowest energy structures found in ion pair calculations for [$C_4C_1Im$][$BF_4$] with the distance N2-B of 0.388 nm using the following methods: **a** and **b**-) Polarizable force field; **c** and **d**-) non-polarizable force field; **e** and **f**-) DFT calculation; **g** and **h**-) semi-empirical calculation. The orientations of the cation are the same than in Figure 3. Selected distances and angle values are given in the bottom of the figure.

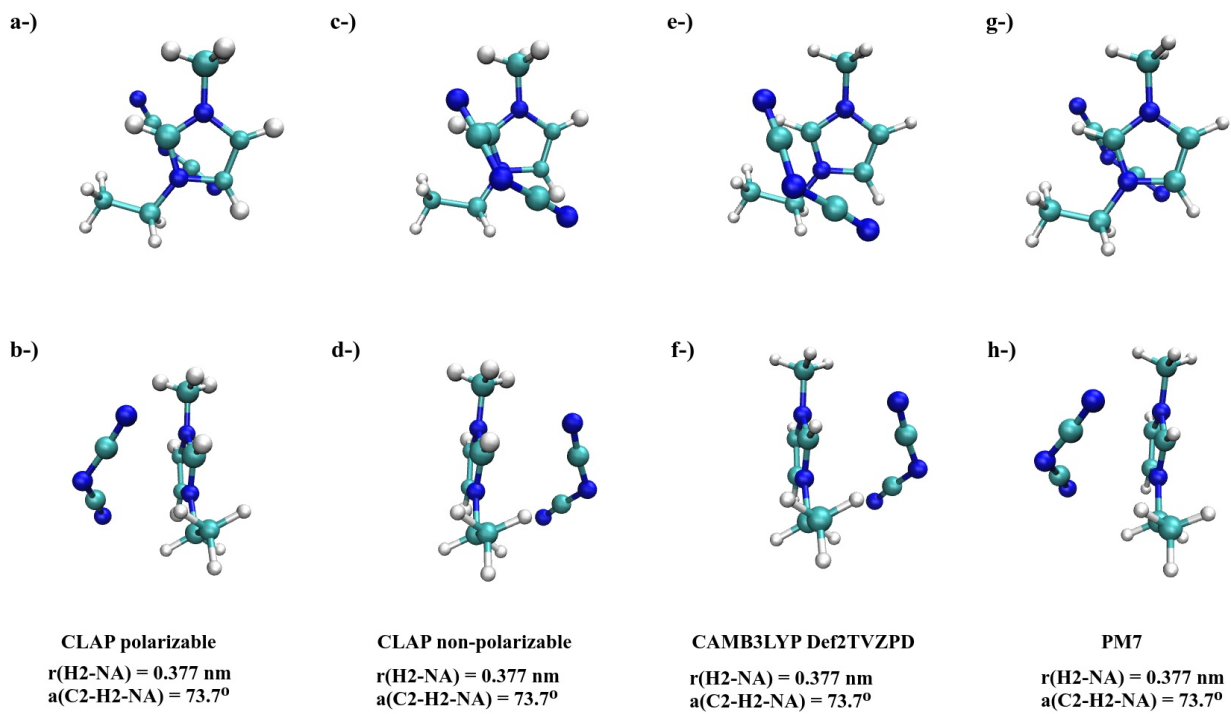

a-)

c-)

e-)

g-)

b-)

d-)

f-)

h-)

**CLAP polarizable**
**r(H2-NA) = 0.377 nm**
**a(C2-H2-NA) = 73.7°**

**CLAP non-polarizable**
**r(H2-NA) = 0.377 nm**
**a(C2-H2-NA) = 73.7°**

**CAMB3LYP Def2TVZPD**
**r(H2-NA) = 0.377 nm**
**a(C2-H2-NA) = 73.7°**

**PM7**
**r(H2-NA) = 0.377 nm**
**a(C2-H2-NA) = 73.7°**

**Figure S15**. Lowest energy structures found in ion pair calculations for $[C_2C_1Im][N(CN)_2]$ with the distance N2-NA of 0.41 nm using the following methods: **a** and **b-)** Polarizable force field; **c** and **d-)** non-polarizable force field; **e** and **f-)** DFT calculation; **g** and **h-)** semi-empirical calculation. The orientations of the cation are the same than in Figure 7. Selected distances and angle values are given in the bottom of the figure.

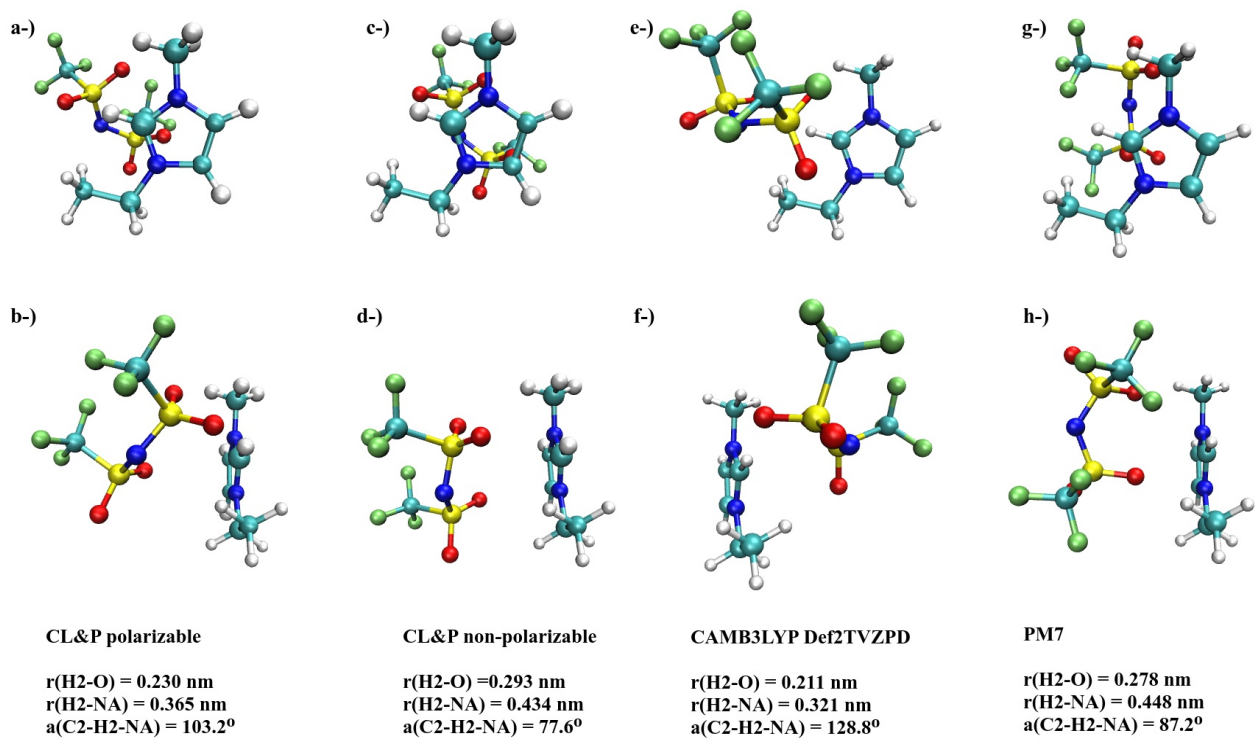

**CL&P polarizable**

**r(H2-O) = 0.230 nm**
**r(H2-NA) = 0.365 nm**
**a(C2-H2-NA) = 103.2°**

**CL&P non-polarizable**

**r(H2-O) =0.293 nm**
**r(H2-NA) = 0.434 nm**
**a(C2-H2-NA) = 77.6°**

**CAMB3LYP Def2TVZPD**

**r(H2-O) = 0.211 nm**
**r(H2-NA) = 0.321 nm**
**a(C2-H2-NA) = 128.8°**

**PM7**

**r(H2-O) = 0.278 nm**
**r(H2-NA) = 0.448 nm**
**a(C2-H2-NA) = 87.2°**

**Figure S16.** Lowest energy structures found in ion pair calculations for $[C_2C_1Im][NTf_2]$ with the distance N2-NA of 0.47 nm and the anion in the C1 conformation using the following methods: **a** and **b-)** Polarizable force field; **c** and **d-)** non-polarizable force field; **e** and **f-)** DFT calculation; *g* and **h-)** semi-empirical calculation. The orientations of the cation are the same than in Figure 12. Selected distances and angle values are given in the bottom of the figure; the oxygen in the distance calculations were taken as the closest one of the H2 atom.

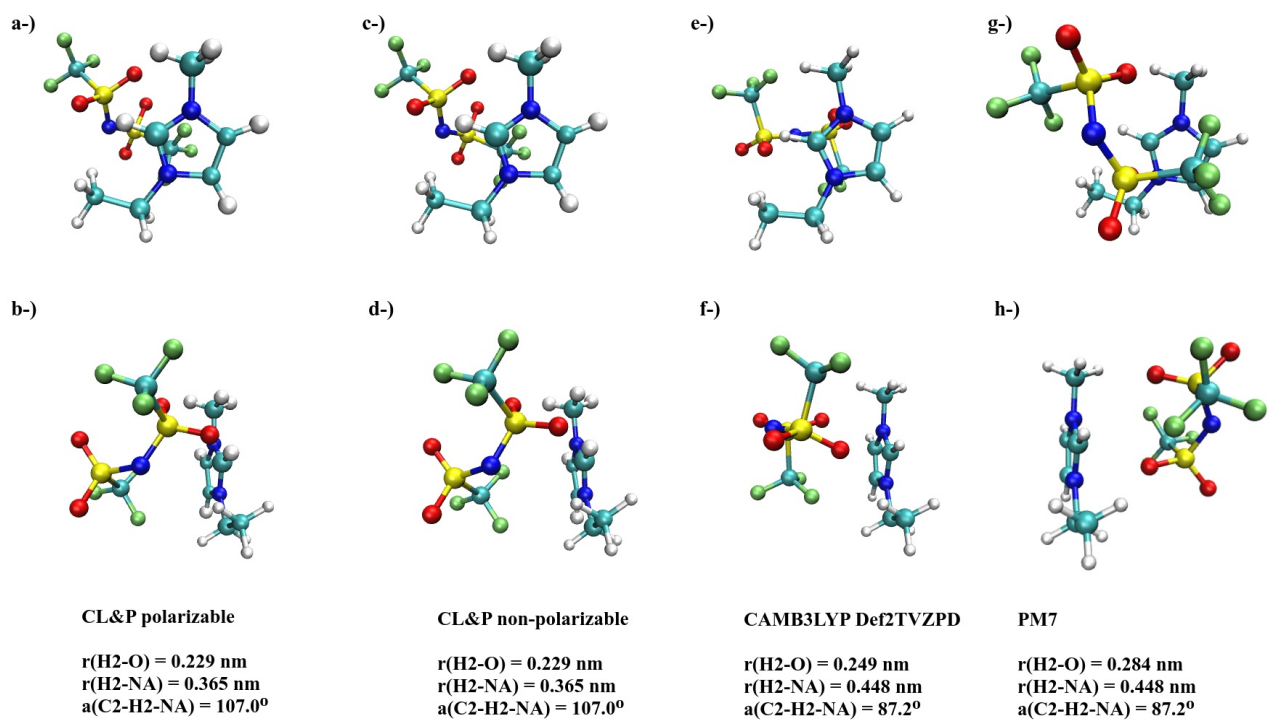

**CL&P polarizable**

**r(H2-O) = 0.229 nm**
**r(H2-NA) = 0.365 nm**
**a(C2-H2-NA) = 107.0º**

**CL&P non-polarizable**

**r(H2-O) = 0.229 nm**
**r(H2-NA) = 0.365 nm**
**a(C2-H2-NA) = 107.0º**

**CAMB3LYP Def2TVZPD**

**r(H2-O) = 0.249 nm**
**r(H2-NA) = 0.448 nm**
**a(C2-H2-NA) = 87.2º**

**PM7**

**r(H2-O) = 0.284 nm**
**r(H2-NA) = 0.448 nm**
**a(C2-H2-NA) = 87.2º**

**Figure S17.** Lowest energy structures found in ion pair calculations for [$C_2C_1Im$][$NTf_2$] with the distance N2-NA of 0.47 nm and the anion in the C2 conformation using the following methods: **a** and **b-)** Polarizable force field; **c** and **d-)** non-polarizable force field; **e** and **f-)** DFT calculation; *g* and **h-)** semi-empirical calculation. The orientations of the cation are the same than in Figure 12. Selected distances and angle values are given in the bottom of the figure; the oxygen in the distance calculations were taken as the closest one of the H2 atom.

**4. Differences in Spatial Distribution Functions (Sdf) Between Polarizable and Non-Polarizable Force Fields**

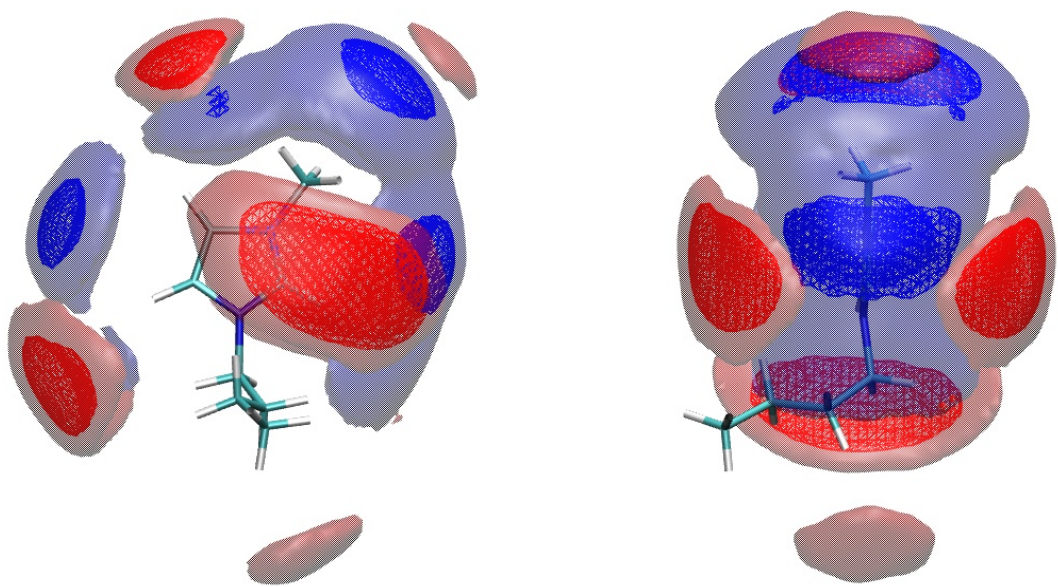

**Figure S18.** Difference between the sfd of B atom of $[BF_4]^-$ around $[C_4C_1Im]^+$ obtained in the simulation with the polarizable force field (Figure 5 **a** and **b**) and the non-polarizable force field (Figure 5 **c** and **d**). The blue surfaces correspond to the regions where the anion density is larger in the polarizable model and the red surfaces correspond to the regions where density is larger in the non-polarizable force field. Isovalues: 5 nm$^{-3}$ (wireframes) and 2 nm$^{-3}$ (translucid surfaces). In the right representation, the HR atom points toward the reader.

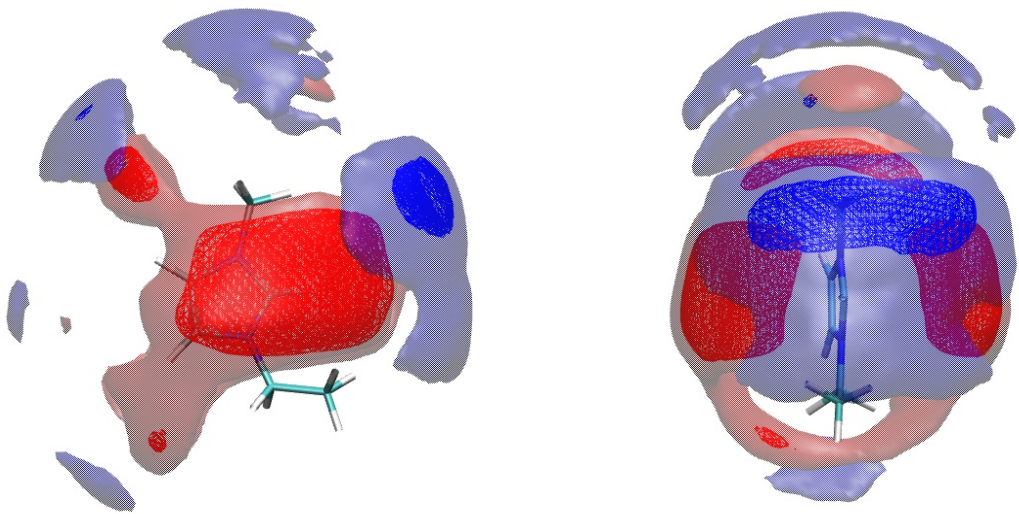

**Figure S19.** Difference between the sfd of NA atom of [N(CN)$_2$]$^-$ around [C$_2$C$_1$Im]$^+$ obtained in the simulation with the polarizable force field (Figure 8 **a** and **b**) and the non-polarizable force field (Figure 8 **c** and **d**). The blue surfaces correspond to the regions where the anion density is larger in the polarizable model and the red surfaces corresponds to the regions where density is larger in the non-polarizable force field. Isovalues: 4 nm$^{-3}$ (wireframes) and 2 nm$^{-3}$ (translucid surfaces). In the right representation, the HR atom points toward the reader.

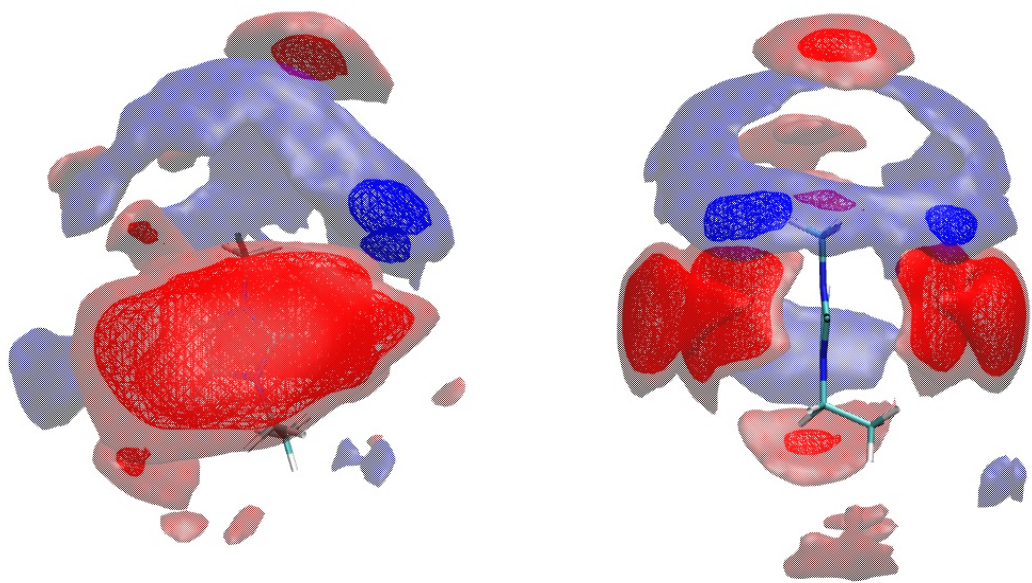

**Figure S20.** Difference between the sfd of NA atom of [NTf$_2$]$^-$ around [C$_2$C$_1$Im]$^+$ cation obtained in the simulation with the polarizable force field (Figure 12 **a** and **b**) and the non-polarizable force field (Figure 12 **c** and **d**). The blue surfaces correspond to the regions where the anion density is larger in the polarizable model and the red surfaces corresponds to the regions where density is larger in the non-polarizable force field. Isovalues: 1.4 nm$^{-3}$ (wireframes) and 0.8 nm$^{-3}$ (translucid surfaces). In the right representation, the HR atom points toward the reader.